\documentclass[12pt,letterpaper]{article}
\usepackage[dvipsnames, table]{xcolor} %Removed usenames option as it warned that it was obselete.
\usepackage[pagebackref=true]{hyperref}
\usepackage[margin=1.0in]{geometry}
\usepackage{cite}

\usepackage{tocloft}
\usepackage{graphicx}
\usepackage{verbatim}
\usepackage[mathscr]{euscript}
\usepackage{slashed}
\usepackage{graphicx}
\usepackage{mathdots}
\usepackage{caption}

\usepackage[labelformat=parens, subrefformat=simple]{subcaption}

\usepackage{enumerate}
\usepackage{bbm}
\usepackage{psfrag}
\usepackage{subfiles}
\usepackage{psfrag}
\usepackage{relsize}
\usepackage{pgfplots}
\pgfplotsset{compat=1.18} %Added because of a warning to hold fixed the compatibility version of pgfplots; see {https://tex.stackexchange.com/questions/81899/what-does-running-in-backwards-compatibility-mode-mean-and-what-should-i-fix-t}
\usepackage{tikzscale}

\usepackage{bbm} 					%%% Blackboard Bold math symbols
\usepackage{slashed} 				%%% Feynman slash
\usepackage{graphicx}				%%% Graphics
\usepackage{subcaption}			%%% Subcaptions for subfigures
\usepackage{psfrag}				%%% Editing EPS files in TeX
\usepackage{tensor}				%%% The \indices command
\usepackage{fouridx}				%%% Arbitrary subscripts/superscripts
\usepackage{bm}					%%% Bold Math symbols
\usepackage{mdframed}				%%% Box for \aside
\usepackage{multirow}				%%% More flexible tables
\usepackage{soul}					%%% Strike-through text
\usepackage{multicol}				%%% Pages with multiple columns

\usepackage{amsmath}
\usepackage{amssymb}
\usepackage{amsthm}

\numberwithin{equation}{section}

\usepackage{mathtools}

\usepackage{feynmf}
\usepackage{marvosym}

\usepackage{import}

\newtheoremstyle{named}{}{}{\itshape}{}{\bfseries}{.}{.5em}{#3}
\theoremstyle{named}

\newtheoremstyle{unnamed}{}{}{\itshape}{}{\bfseries}{.}{.5em}{\thmname{#1}\thmnumber{ #2}}
\theoremstyle{unnamed}

\newtheoremstyle{namedandnumbered}{}{}{\itshape}{}{\bfseries}{.}{.5em}{\thmname{#1}\thmnumber{ #2} (\thmnote{#3})}
\theoremstyle{namedandnumbered}

\theoremstyle{definition}

\usepackage{diagbox}

\newcommand*\xoverline[2][0.75]{%
    \sbox{\myboxA}{$\m@th#2$}%
    \setbox\myboxB\null% Phantom box
    \ht\myboxB=\ht\myboxA%
    \dp\myboxB=\dp\myboxA%
    \wd\myboxB=#1\wd\myboxA% Scale phantom
    \sbox\myboxB{$\m@th\overline{\copy\myboxB}$}%  Overlined phantom
    \setlength\mylenA{\the\wd\myboxA}%   calc width diff
    \addtolength\mylenA{-\the\wd\myboxB}%
    \ifdim\wd\myboxB<\wd\myboxA%
       \rlap{\hskip 0.5\mylenA\usebox\myboxB}{\usebox\myboxA}%
    \else
        \hskip -0.5\mylenA\rlap{\usebox\myboxA}{\hskip 0.5\mylenA\usebox\myboxB}%
    \fi}
\makeatother

\newcommand{\ZZ}{\mathbb{Z}}

\newcommand{\cE}{\mathcal{E}}
\newcommand{\cF}{\mathcal{F}}

\newcommand{\cK}{\mathcal{K}}

\newcommand{\cM}{\mathcal{M}}

\newcommand{\Mpl}{M_{\textrm{Pl}}}

\newcommand{\cO}{\mathcal{O}}

\newcommand{\cS}{\mathcal{S}}

\newcommand{\cV}{\mathcal{V}}

\newcommand{\tr}{\,\mathrm{tr}}

\definecolor{cobalt}{RGB}{44, 98, 120}
\definecolor{celadon}{rgb}{0.67, 0.88, 0.69}
\definecolor{dm}{cmyk}{.20, 0, .30, 0}
\definecolor{burgundy}{rgb}{0.5, 0.0, 0.13}
\definecolor{plotBlue}{RGB}{94, 130, 181}

\newcommand{\rmd}{\textrm{d}}
\def\be{\begin{equation}}
\def\ee{\end{equation}}
\def\bea{\begin{eqnarray}}
\def\eea{\end{eqnarray}}

\hypersetup{
  colorlinks,
  citecolor=Violet,
  linkcolor=cobalt,
  urlcolor=Blue}

\newif\iffastcompile

\fastcompilefalse

\iffastcompile
\newcommand{\js}[1]{}
\newcommand{\jsi}[1]{}
\newcommand{\cl}[1]{}
\newcommand{\lm}[1]{}
\else
\newcommand{\js}[1]{\todo[color=cobalt!30,size=\scriptsize, bordercolor=cobalt!30]{JS: #1}}
\newcommand{\jsi}[1]{\todo[color=cobalt!30,size=\scriptsize, bordercolor=cobalt!30, inline]{JS: #1}}
\newcommand{\cl}[1]{\todo[color=burgundy!30, size=\scriptsize, bordercolor=burgundy!30]{CL: #1}}
\newcommand{\lm}[1]{\todo[color=dm!90, size=\scriptsize, bordercolor=dm!90]{LM: #1}}
\fi

\ProvideTextCommandDefault{\Dbar}{%
\leavevmode\lower.5ex\rlap{\hskip-.07em\accent"16}D%
}

\newcommand{\secref}[1]{\S\ref{#1}}

\newcommand{\orthogcomplement}{\bot}

\renewcommand{\ker}{\operatorname{Ker}}
\newcommand{\deriv}{\mathrm{d}}
\newcommand{\brap}[1]{{\left( {#1} \right)}}
\newcommand{\bras}[1]{{\left[ {#1} \right]}}
\newcommand{\brac}[1]{{\left\{ {#1} \right\}}}

\newcommand{\brav}[1]{{\left\vert {#1} \right\vert}}
\newcommand{\bravv}[1]{{\left\Vert {#1} \right\Vert}}
\newcommand{\bbR}{\mathbb{R}}

\newcommand{\bbZ}{\mathbb{Z}}
\graphicspath{{./fig}} %Tell LaTeX where to find images for {\includegraphics}'s.

\newcommand{\cEperp}{\cE_{\text{perp}}}
\newcommand{\cEker}{\cE_{\text{ker}}}
\newcommand{\cElimdir}{\cE_{0}}

\newcommand{\cMtau}{\cM_{\text{parallel}}}
\newcommand{\cMkerorth}{\cM_{\text{ker}^\bot}}
\newcommand{\cMlight}{\cM_{\text{L}}}
\newcommand{\tttunderscore}{\_\allowbreak }
\newcommand{\CYTools}{\texttt{CYTools}}
\newcommand{\Vol}{{\cV}}
\newcommand{\tildeq}{{\tilde q}}
\newcommand{\metric}{\mathfrak{g}}
\newcommand{\metricInvIndexed}{\mathfrak{g}}
\newcommand{\metricInvNoIndex}{{\mathfrak{g}^{-1}}}
\newcommand{\rhoAx}[2]{{\rho^{\brap{\text{ax}}}_{C_{#1};#2}}}
\newcommand{\rhoInst}[2]{{\rho^{\brap{\text{inst}}}_{C_{#1};#2}}}
\newcommand{\alphaAx}[2]{{\alpha^{\brap{\text{ax}}}_{C_{#1};#2}}}
\newcommand{\alphaInst}[2]{{\alpha^{\brap{\text{inst}}}_{C_{#1};#2}}}
\newcommand{\alphaBoth}[1]{{\alpha_{C_{#1}}}}

\begin{document}
	\newcommand{\main}{.}
\begin{titlepage}

\setcounter{page}{1} \baselineskip=15.5pt \thispagestyle{empty}

\bigskip\

\bigskip\

\vspace{2cm}
\begin{center}
%{%\fontsize{20}{28}
{\LARGE \bfseries Quantum Gravity Cutoff from Axions:\\ A Type IIB Landscape Study}

 \end{center}
\vspace{0.5cm}
%\vspace{0.25cm}

\begin{center}
{\fontsize{14}{30}\selectfont Matthew Reece$^a$, Tom Rudelius$^b$, Christopher Tudball$^b$}
\end{center}

\begin{center}

\vspace{0.25 cm}
\textsl{$^a$Leinweber Institute for Theoretical Physics, Department of Physics, Harvard University, Cambridge, MA, 02138, USA}\\
				\textsl{$^b$Department of Mathematical Sciences, Durham University, Durham, DH1 3LE, UK}\\

\vspace{1cm}mreece@g.harvard.edu, thomas.w.rudelius@durham.ac.uk, christopher.a.tudball@durham.ac.uk

\end{center}

\vspace{1cm}
\noindent 
Extra-dimensional axions have coupling strength related to fundamental, ultraviolet physics. It has been proposed that the properties of such axions imply a bound on the quantum gravity cutoff: $\Lambda_\textsc{QG} \lesssim 2\pi \sqrt{S} f$, where $f$ is the axion decay constant and $S$ is the instanton action. In the context of weakly-coupled string theory, we identify $\Lambda_\textsc{QG}$ with the string scale $M_s$. In this paper, we carry out a quantitative study of this bound on the string scale in the context of Calabi-Yau compactifications of Type IIB string theory, considering both $C_2$ and $C_4$ axions. We show, both analytically and numerically, that the bound holds even near boundaries of the K\"ahler moduli space, including those where the co-scaling relationship for axion strings fails. This evidence bolsters previous arguments, based on naturalness and on unitarity, that the bound is a general feature of extra-dimensional axions in quantum gravity.

 \vspace{1.1cm}
% \hrule

\bigskip
\noindent\today

\end{titlepage}
\setcounter{tocdepth}{2}
\tableofcontents

\section{Introduction}\label{INTRO}

The QCD axion potentially solves the Strong CP problem~\cite{Peccei:1977ur,Peccei:1977hh,Weinberg:1977ma,Wilczek:1977pj} and provides a natural dark matter candidate~\cite{Preskill:1982cy, Dine:1982ah, Abbott:1982af}. Models in which the axion arises from a higher dimensional gauge field are particularly compelling, from the bottom-up perspective as a natural solution of the Strong CP problem and from the top-down perspective as a prediction of string theory~\cite{Witten:1984dg, Choi:1985je, Barr:1985hk, Dine:1986bg, Banks:1996ea, Choi:2003wr, Conlon:2006tq, Svrcek:2006yi, Arvanitaki:2009fg, Acharya:2010zx, Cicoli:2012sz, Demirtas:2018akl, Demirtas:2021gsq, Gendler:2023kjt, Fallon:2025lvn}; see~\cite{Reece:2023czb, Craig:2024dnl, Reece:2025thc} for recent expositions. 

In a theory of a single axion $\theta \cong \theta + 2\pi$, the most important parameters are the decay constant $f$ and the Chern-Simons levels $k_F$ and $k_G$ for the axion's coupling to photons and gluons:
\begin{equation}
S = \int \left(-\frac{1}{2} f^2 |\rmd\theta|^2 - \frac{k_G}{8\pi^2} \theta\, \mathrm{tr}(G \wedge G) - \frac{k_F}{8\pi^2} \theta F \wedge F \right).
\end{equation}
The level $k_G$ must be an integer, while $k_F$ satisfies a slightly more complicated quantization condition~\cite{Reece:2023iqn, Choi:2023pdp, Cordova:2023her}. They are both typically order-one.
Most experimental searches for the axion target the axion-photon coupling
\begin{equation}
g_{a\gamma\gamma}  = \frac{\alpha}{2\pi f} \left(2 k_F - 1.92(4) k_G\right),
\end{equation}
which depends not only on the Chern-Simons coupling $k_F$ to photons but also on $k_G$ through QCD strong dynamics~\cite{Kaplan:1985dv,Srednicki:1985xd,Georgi:1986df}. For the quantitative value of the constant multiplying $k_G$ we have quoted \cite{GrillidiCortona:2015jxo}, although there is an active effort underway to improve on these computations~\cite{Lu:2020rhp, Gao:2022xqz, Meggiolaro:2025yiu} including a recent effort to compute it using Lattice QCD~\cite{Brandt:2026wir}. A discovery of a QCD axion would also shed light on the axion decay constant $f$ via the relationship
\begin{equation}
m_a \approx 5.7\,\mu\mathrm{eV} \left(\frac{10^{12}\,\mathrm{GeV}}{f/k_G}\right).
\end{equation}
Assuming that $k_F$ and $k_G$ are order-one numbers, as expected in typical models, an experimental discovery of an axion would tell us about the energy scale $f$. For a QCD axion, the measurement of both $g_{a\gamma\gamma}$ and $m_a$ would provide an important cross-check on this determination.

Measuring the axion decay constant $f$ would provide an important clue about high-energy physics. The axion-photon and axion-gluon couplings are non-renormalizable and thus lead to amplitudes that grow with energy. For a QCD axion, a perturbative unitarity argument based on scattering the color-singlet two-gluon state $\frac{1}{N_c^2 - 1} \sum_a |g^a g^a\rangle$~\cite{Brivio:2021fog} implies that the effective field theory breaks down at a scale
\begin{equation} \label{eq:unitarityargument}
\Lambda_\textsc{EFT} \leq S \frac{f}{k_G} \sqrt{\frac{\pi}{N_c^2 - 1}},
\end{equation}
where 
\begin{equation}
S  = \frac{8\pi^2}{g^2}
\end{equation}
is the instanton action. This implies that any QCD axion model has new physics at an energy scale no more than about $100 f/k_G$. In a conventional 4d axion model, the scattering amplitude is unitarized by new fermions that are integrated out perturbatively at one loop. These fermions necessarily have a mass that is less than or around the scale $f$. 

For an extra-dimensional axion model, on the other hand, the 4d Chern-Simons term descends from a higher-dimensional Chern-Simons term. We then expect that the Kaluza-Klein scale lies below the energy~\eqref{eq:unitarityargument}, and the physics that unitarizes scattering amplitudes is fundamental, i.e., it is associated with the quantum gravity UV completion. This is particularly clear for the case of a {\em fundamental axion}, i.e., one in which the core of an axion string lies at infinite distance in field space, so there is no sense in which a Peccei-Quinn symmetry can be restored~\cite{Reece:2018zvv}. In this context, an even stronger bound has been proposed~\cite{Reece:2025thc} (and further studied in~\cite{Benabou:2025kgx,Reece:2025zva}). We refer to this as the {\em Quantum Gravity Cutoff from Axions}:
\begin{equation} \label{eq:LambdaQGbound}
\Lambda_\textsc{QG} \lesssim 2\pi \sqrt{S} f.
\end{equation}
Here $\Lambda_\textsc{QG}$ is the quantum gravity scale, or species scale~\cite{Veneziano:2001ah}, at which local QFT breaks down. In string theory, we will identify this scale with the string scale $M_s$. In practice, then, we are mostly interested in investigating the bound
\begin{equation}
M_s \lesssim 2\pi \sqrt{S} f \,. 
\label{stringscalebound}
\end{equation}
We will refer to this special case of the Quantum Gravity Cutoff from Axions as the {\em String Scale Cutoff from Axions} or, in the remainder of this paper where unambiguous, simply ``the string scale bound.''

A few different arguments have been given for the bound~\eqref{eq:LambdaQGbound}, none of which is fully rigorous:
\begin{itemize}
\item {\em Naturalness.} The axion propagator gets loop corrections from an entire tower of gluon Kaluza-Klein modes. The $\theta g_n g_n$ three-point amplitude is expected to be of order $g^2/(8\pi^2 f)$ for any KK mode $g_n$. The number of KK modes below the cutoff is proportional to the volume of the extra dimensional cycle, which is in turn proportional to the gauge kinetic factor $1/g^2$. Summing up these loop corrections and requiring that they not dominate over the tree-level propagator leads to a constraint that is parametrically of the form~\eqref{eq:LambdaQGbound}~\cite{Reece:2025thc} (see also~\cite{Seo:2024zzs}).

\item {\em Perturbative Unitarity.} The perturbative unitarity constraint~\eqref{eq:unitarityargument} can be improved by scattering states that sum over KK mode number, $\frac{1}{\sqrt{(N_c^2 - 1)N_\textsc{KK}}} \sum_a \sum_n |g_n^a g_n^a\rangle$~\cite{Benabou:2025kgx}. This again parametrically leads to a bound of the form~\eqref{eq:LambdaQGbound}.

\item {\em Electric-Magnetic Co-Scaling.} In many theories, electrically and magnetically charged objects have tensions related by $T_E / T_M \sim g^2/(2\pi)$, with $g$ the electric coupling~\cite{Reece:2025thc, Reece:2025zva}. In the axion case, the electric coupling is $1/f$; the electrically charged object is an instanton, with tension $T_E$ equal to the action $S$; and the magnetically charged object is an axion string with tension $\mathcal{T}$. In that case, the co-scaling relation is
\begin{equation} \label{eq:axioncoscaling}
    \mathcal{T} \sim 2\pi f^2 S.
\end{equation}
For extra-dimensional axions, the axion string is a fundamental object (string or brane) wrapping a cycle that should be large in units of the species scale for geometric control~\cite{Martucci:2024trp}. (In particular, in cases where co-scaling applies, the axion string is an EFT string in the language of~\cite{Lanza:2020qmt,Lanza:2022zyg}.) In this case,~\eqref{eq:axioncoscaling} together with $2\pi \mathcal{T} \gtrsim \Lambda_\textsc{QG}^2$ leads to the bound~\eqref{eq:LambdaQGbound}~\cite{Reece:2025thc}.

\item {\em Examples in String Theory.} The bound~\eqref{stringscalebound} was explicitly checked in a few examples in~\cite{Reece:2025thc}, all of which exhibited electric-magnetic co-scaling. Subsequently, the bound was checked at the tip of the stretched K\"ahler cone (defined in~\cite{Long:2016jvd,Demirtas:2018akl}) in a large scan of Type IIB Calabi-Yau compactification examples~\cite{Benabou:2025kgx} (see also~\cite{Benabou:2026jtv}), using \CYTools{}~\cite{Demirtas:2022hqf}. Related estimates also appear in \S3.3 of~\cite{Cheng:2025ggf}, which found that in the IIB landscape
\begin{equation}
    M_s \approx O(1) S^{1/4} f \kappa_{\mathrm{max}}^{1/3}
\end{equation}
is a good approximation (especially at large $h^{1,1}$), where $\kappa_{\mathrm{max}}$ is the self-triple-inter\-sec\-tion number for the divisor of largest volume. When $S \gg 1$, this is compatible with~\eqref{stringscalebound}.

\end{itemize}

This earlier work shows that~\eqref{eq:LambdaQGbound} holds in many extra-dimensional axion theories. However, it is not fully satisfying. The unitarity and naturalness arguments signal that something goes wrong at an energy scale of order $\sqrt{S} f$, but they rely on rough estimates of the coupling of the axion to KK modes. Electric-magnetic co-scaling is a common feature of quantum gravity theories, but it is known to fail in certain cases (e.g., near a flop transition), although in one such example it was checked that~\eqref{stringscalebound} still holds~\cite{Reece:2025zva}. Similarly, the scan over Calabi-Yau manifolds conducted in~\cite{Benabou:2025kgx} focused on the tip of the stretched K\"ahler cone, potentially missing qualitatively different behavior in other regions of moduli space. Finally, many of the arguments (like the unitarity discussion above) were developed for a single axion, and it is important to clarify what $f$ and $S$ mean in a multi-axion context.

In this paper, we carry out a further study of examples in Type IIB string theory, with a special focus on corners of the parameter space where co-scaling is known to fail, and to different possible choices of $f$ and $S$ in theories with multiple axions. We will see, both analytically and numerically, that~\eqref{stringscalebound} continues to hold.

In \S\ref{sec:setup}, we explain how we generalize~\eqref{stringscalebound} to cases with multiple axions. We fix our conventions and outline the calculations we will perform for $C_2$ and $C_4$ axions in Type IIB string theory. In \S\ref{sec.analytics}, we present analytic arguments that $M_s$ does not grow large relative to $\sqrt{S} f$ near boundaries of the K\"ahler cone. In \S\ref{sec:numerics}, we present numerical results from scans using \CYTools{}, confirming that the bound~\eqref{stringscalebound} is robust both at the tip of the K\"ahler cone and near facets. Finally, we offer concluding remarks in \S\ref{sec:conclusions}.

\section{Setup}
\label{sec:setup}

    Consider a theory with a single axion $\theta$, with period $\theta \simeq \theta+2\pi$ and decay constant $f$. The contribution of its kinetic term to the action is
    \begin{equation}
        \int \sqrt{\brav{g}} \deriv^4x \brap{ -\frac{f^2}2 \partial_\mu\theta\partial^\mu\theta}\,.
    \end{equation}
    Suppose there is also an instanton of unit charge under the axion with instanton action $S$.
    This instanton experiences a gauge coupling of $f^{-1}$.

    We wish to verify \eqref{stringscalebound}, i.e., to check that the string scale $M_s$ is no more than an order-one number larger than $2\pi\sqrt{S}f$. This is equivalent to checking that $\rho$ is order-one or smaller, where
    \begin{equation}\label{eqn.rho.single.axion}
		\rho:=\frac{M_s}{2\pi\sqrt{S}f}\,.
	\end{equation}

\subsection{The challenge of the multi-axion case}\label{ss.challenge}    

    The generalization of \eqref{eqn.rho.single.axion} to theories with multiple axions is subtle, and there is more than one way to define the quantity $\rho$.

    The most experimentally relevant quantities are the mass of a QCD axion, $m_a$, and its coupling to photons, $g_{a\gamma\gamma}$. If there are multiple axions, what will be measured in any experiment is essentially these properties for a kinetic and mass eigenstate. It is always possible to simultaneously diagonalize the kinetic and mass terms, but this obscures the underlying periodicity properties of the axions and does not lend itself well to analytic studies. Furthermore, although the kinetic matrix of the axions we consider is determined in string theory by the geometry of the underlying Calabi-Yau manifold, the potential for the axions is not. For a QCD axion, we know that the axion potential depends on the details of confinement and the spectrum of quark masses~\cite{DiVecchia:1980yfw}. The same will, in general, be true in a string compactification: to compute the potential for all of the axion fields, we must specify the full set of D-branes in the theory and the chiral matter fields they give rise to, understand the scale of SUSY breaking, and so on. Past studies that have numerically computed an axion mass spectrum (see, e.g.,~\cite{Demirtas:2018akl, Halverson:2019kna,Demirtas:2021gsq,Mehta:2021pwf, Gendler:2023kjt,Benabou:2025kgx}) must make a number of assumptions to do so. Despite these challenges, we can make some plausible assumptions to argue that convenient, calculable quantities are generally a good proxy for the underlying physical couplings of the axion. Our discussion is partly based on that of~\cite{Gendler:2023kjt}.

    To begin, consider an integer basis of $n$ axions, by which we mean that there are $n$ independent gauge redundancies $\theta^i \cong \theta^i + 2\pi$ (for $i = 1, \ldots n$). The action allows for an arbitrary kinetic mixing among these axions:
    \begin{equation} \label{eq:generalaction}
      S = \int \sqrt{|g|}\rmd^4x\,\left(-\frac{1}{2} \kappa_{ij} \partial_\mu \theta^i \partial^\mu \theta^j - V(\{\theta\})\right).  
    \end{equation}
    We will assume that the axion potential is a sum of $n$ terms giving rise to masses for $n$ independent linear combinations of the axions:
    \begin{equation} \label{eq:Vintegerbasis}
        V(\{\theta\}) \approx \sum_{i = 1}^n V_i(\textstyle\sum_{j=1}^n c^{i}_{\,j} \theta^j), \quad c^{i}_{\,j} \in \ZZ, \quad \det c \neq 0,
    \end{equation}
    where each $V_i$ is a periodic function and we can assume, without loss of generality, that the greatest common divisor of the set of $c^{i}_{\,j}$ for fixed $i$ is 1.
    The main assumption that we are making here is that further terms in the potential can be treated as small perturbations. This is likely to be true if the potentials are all generated from nonperturbative effects: the characteristic magnitude $|V_i|$ of a term in the potential is exponentially small compared to the cutoff scale. It is natural then to have hierarchically separated scales for the different terms in the potential, and we will assume an ordering $|V_1| \gg |V_2| \gg \cdots |V_n| \gg |V_{n+1}| \gg \cdots$.  To further simplify, without loss of generality, one can go to an integer basis where $V_1$ gives a mass to $\theta^1$, $V_2$ to $\theta^2$, and so on. This arises from a $\mathrm{GL}(n,\ZZ)$ transformation to the {\em Smith normal form} of the mass matrix~\cite{smith1862systems} (see, e.g.,~\cite{Fraser:2019ojt,Choi:2020rgn,Gendler:2023hwg} for past use in the axion mixing context). Thus, we are free to assume:
    \begin{equation}
        V(\{\theta\}) \approx \sum_{i=1}^n V_i(\theta^i).
    \end{equation}
    At this stage, the kinetic mixing matrix is still general. Past work has indicated that in string compactifications, the off-diagonal terms of $\kappa$ are usually very small in this  basis~\cite{Halverson:2019cmy,Gendler:2023kjt}. However, we will forge ahead: we can make a useful statement even without assuming that kinetic mixing among the $\theta^i$ is suppressed.

    The main idea now is to exploit the hierarchical nature of the $|V_i|$. Let us illustrate the idea by focusing first on the two heaviest fields. We can eliminate kinetic mixing among $\theta^1$ and $\theta^2$ by carrying out a field redefinition of the form $\theta^2 \mapsto \theta^2 + \xi \theta^1$ for some constant $\xi$. The term $V_1(\theta^1)$ is unchanged, and dominant over $V_2$, so it stabilizes $\theta^1$ and determines the mass of the physical mode. In the next term, now of the form $V_2(\theta^2 + \xi \theta^1)$, we can approximately set $\theta^1$ to its constant VEV determined by $V_1$, and {\em still} view the potential as a function of (the redefined) $\theta^2$ only, which is then stabilized by the $V_2$ term. For the general $n$-field case, the procedure is to carry out a {\em UDU decomposition} of the kinetic matrix~\cite{benoit1924note, watkins2004fundamentals}. Because $\kappa$ is a positive definite symmetric matrix, we have
    \begin{equation} \label{eq:UDU}
        \kappa = U D U^\top,
    \end{equation}
    where $U$ is an upper triangular matrix with only 1s on the diagonal, and $D$ is a real diagonal matrix with all diagonal elements positive. This corresponds to a basis change to the fields
    \begin{equation}
        \varphi^i = (U^\top)^i_{\,j} \theta^j.
    \end{equation}
    $U^\top$ is a lower triangular matrix, so its inverse matrix $L = (U^\top)^{-1}$ is as well. We then have
    \begin{equation}
        \theta^i = L^i_{\,j} \varphi^j = \varphi^i + \sum_{j < i} L^i_{\,j} \varphi^j.
    \end{equation}
    This gives rise to a potential with the structure
    \begin{equation}
        V(\varphi) = V_1(\varphi^1) + V_2(\varphi^2 + L^2_{\,1}\varphi^1) + V_3(\varphi^3 + L^3_{\,2} \varphi^2 + L^3_{\,1} \varphi^1) + \cdots.
    \end{equation}
    When we first integrate out $\varphi^1$ using $V_1$, the next term $V_2$ is effectively a function only of $\varphi^2$, which we then integrate out so that $V_3$ is effectively a function only of $\varphi^3$, and so on. The change of basis leaves intact the conclusion that $V_i$ gives a mass to only the $i$th axion, but now the axions have a diagonal kinetic matrix.
    
    The diagonal kinetic matrix $D$ in the $\varphi$ basis determines the ``effective decay constants''  that will be measured experimentally  (up to $|V_i|/|V_j|$ suppressed effects with $j > i$):
    \begin{equation}
        D = \begin{pmatrix} f_{\mathrm{eff;}1}^2 & & \\ & \ddots & \\ & & f_{\mathrm{eff};n}^2 \end{pmatrix}.
    \end{equation}
The question, then, is to what extent the effective decay constants $f_{\mathrm{eff};i}$ can differ significantly from the periodicities $f_i$ defined by 
\begin{equation}
f_i^2 := \kappa_{ii}\,,~ ~~~\text{(no sum over $i$),}
\label{fdiag}
\end{equation}
which are the quantities that are more straightforwardly computed from geometric data. In the bulk of this paper, we will present evidence for a bound of the form $M_s \lesssim 2\pi\sqrt{S_i} f_i$. Clearly, this bound can be applied to the measured quantity $f_{\mathrm{eff};i}$ whenever  $f_{\mathrm{eff};i} \geq f_i$, whereas if $f_{\mathrm{eff};i} \ll f_i$ then $M_s \lesssim 2\pi\sqrt{S_i} f_{\mathrm{eff};i}$ may fail to hold. 
From~\eqref{eq:UDU} we have
\begin{equation}
    \det \kappa = \det D.
\end{equation}
In fact, we can make a stronger statement: any eigenvector of $D$ with eigenvalue zero determines an eigenvector of $\kappa$ with eigenvalue zero. In particular, suppose that $f^2_{\mathrm{eff};i} = 0$. Then $v^j = \delta^j_i$ is a zero eigenvector of $D$, so there is a zero eigenvector of $\kappa$ given by $q^k = L^k_{\,j} v^j = L^k_{\,i}$. As we will discuss further in \S\ref{sec.analytics}, in the context of the Calabi-Yau geometries studied in this work, one can argue that the kernel of $\kappa$ is generated by integral vectors.

This establishes that there is an integral linear combination of our basis axions with decay constant $f \to 0$ when $f_{{\rm eff};i} \to 0$. We should further exclude the possibility that the effective decay constant goes to zero at a faster rate, with $f_{{\rm eff};i} \ll f$, which could lead to parametric violation of~\eqref{stringscalebound}. However, because $\det \kappa = \det D$, this would require that the product of the remaining eigenvalues of $D$ goes to infinity. Here another constraint is useful: because the entries of $D$ are positive, a straightforward computation shows that $\tr\, D \leq \tr\, \kappa$ (with equality only when $\kappa$ is already diagonal). This forbids the case where some eigenvalue of $D$ goes to infinity without a corresponding diverging eigenvalue of $\kappa$.

Summing up: when kinetic mixing effects are small, the propagating axion eigenstates are approximately  integer linear combinations of $2\pi$-periodic axions $\theta$, as in~\eqref{eq:Vintegerbasis}. When kinetic mixing effects are most dramatic, they single out a direction with a small effective decay constant, and this direction is again approximately an integer linear combination of the $2\pi$-periodic axions $\theta$. This gives us confidence that studying the decay constants of axions that are defined by integer linear combinations is likely to cover all of the cases of interest. We will also take special care to understand the physics near boundaries of the K\"ahler cone, where the effective decay constants may become small. To further support our conclusion, we will study multiple possible proxies for effective axion decay constants.

\subsection{Defining proxies for the multi-axion case}\label{ss.proxies}   

We work in an integer basis of $n$ axions with periodicities~\eqref{fdiag}. We will define two proxies for the physically observable axion couplings, which are basis independent but depend on a choice of charge, either in the (electric) charge lattice of instantons or the (magnetic) charge lattice of axion strings. 
    
First, let us consider a general axion string with monodromy $\theta^i \rightarrow \theta^i + 2 \pi {P}^i$. The charge of this string is given by
\begin{equation}\label{eq:f-axion-string}
f_{P} := \sqrt{{P}^i \kappa_{ij} {P}^j}\,.
\end{equation}
Note that this reduces to \eqref{fdiag} in the simple case ${P}^i = \delta^i_j$. It also has the form of the decay constants in the basis selected by the Smith normal form explained in the previous subsection.
From here, we may define a generalization of \eqref{eqn.rho.single.axion} as:
    \begin{equation}
        {\rho^{\brap{\text{ax}}}_{P,Q}}
        := \frac{M_s}{2\pi \sqrt{S_Q}f_{P}}\,,
        \label{rhoax}
    \end{equation}
    where $S_Q$ is the action of an instanton with nonzero charge under the relevant axion with axion string ${P}^i$, i.e., $Q_i{P}^i \neq 0$.

    There is however another possible generalization of \eqref{eqn.rho.single.axion}.
    While for a single axion the charge of the instanton of integral charge $Q=1$ is given by $f^{-1}=\sqrt{Q \cdot f^{-2} \cdot Q}$, in the case of multiple axions the charge of the instanton is instead given by
    \begin{equation}\label{eq:f-instanton}
        f_Q^{-1} := \sqrt{Q_i \kappa^{ij} Q_j}\,,
    \end{equation}
    where $\kappa^{ij}$ is the inverse of $\kappa_{ij}$.  With this, we may alternatively define:
    \begin{equation}
        {\rho^{\brap{\text{inst}}}_{Q}}:=\frac{M_s}{2\pi \sqrt{S_Q}f_Q}\,.
        \label{rhoinst}
    \end{equation}
    The decay constant $f_Q$ is precisely the quantity bounded by the axion weak gravity conjecture \cite{Arkanihamed:2006dz,rudelius:2015xta, Brown:2015iha, Harlow:2022ich, DiUbaldo:2026rly, Maldacena:2026jqd, Etheredge:2026rio}.

    In what follows, we shall consider both~\eqref{rhoax} and~\eqref{rhoinst} as possible generalizations of \eqref{eqn.rho.single.axion}, for both $C_2$ and $C_4$ axions in compactifications of Type IIB string theory on Calabi-Yau threefolds.
    In this context, we shall give analytic arguments that $\rho$ remains finite as we approach a boundary of moduli space. Additionally, we will perform numerical scans with \CYTools{}~\cite{Demirtas:2022hqf} to check that $\rho \lesssim 1$ near boundaries and at the tip of the stretched K\"ahler cone~\cite{Long:2016jvd, Demirtas:2018akl}.
    
    In the end, we will find that both definitions of $\rho$ give us similar results, and in particular neither definition of $\rho$ ever grows much larger than $1$. Together with the arguments in the preceding subsection, this gives us confidence that our results are robust and should apply to an experimentally measurable definition of $f$.

\subsection{Conventions}    
    We now explain our conventions, which (up to factors of $2 \pi$) are based on those in \cite{Grimm:2007hs}, making contact with quantities computable by \CYTools{} \cite{Demirtas:2022hqf}.

    To begin, we assume that all curve and divisor volumes are sufficiently large that we can ignore instanton effects. Then, we let $\cF=\frac16 C_{ijk}t^it^jt^k$, for triple intersection numbers $C_{ijk}$ and K\"ahler coordinates $t^i$, where $t^i$ measures the volume of a 2-cycle in string units. We define $\cF_{ij}=\partial_{t^i}\partial_{t^j}\cF$ etc.,
    and $\cF^{ij}$ as the inverse of $\cF_{ij}$, i.e., $\cF_{ik}\cF^{kj} = \cF^{jk}\cF_{ki} = \delta_i^j$.
    We further define
    \begin{align}
        \Vol = \cF\,,~~~~
        \tau_i = \cF_i\,,~~~~
        \metric_{ij} = \frac{\cF_i\cF_j - \cF_{ij}\cF}{\cF}\,,
    \end{align}
    and $\metricInvNoIndex$, with indices $\bras{\metricInvNoIndex}^{ij}=\metricInvIndexed^{ij}$, the inverse of $\metric$, i.e.~$\metricInvIndexed^{ij}\metric_{jk}=\delta^i_k$.
    With this, $\Vol$, $\tau$, $4\Vol\metric_{ij}$ are generated by the \CYTools{} functions \texttt{compute\tttunderscore{}cy\tttunderscore{}volume}, \texttt{compute\tttunderscore{}divisor\tttunderscore{}volumes}, and \texttt{compute\tttunderscore{}inverse\tttunderscore{}kahler\tttunderscore{}metric} functions respectively.
    
	We also use conventions
	\begin{equation}
		m_s = \frac1{\ell_s} = \frac1{2\pi}M_s
		= \sqrt{\frac{g_s^2}{4\pi\Vol}}\Mpl
		\,.
	\end{equation}
    Here, $M_s$ determines the mass scale of oscillator modes of the fundamental string, which we identify with the quantum gravity cutoff (species scale) $\Lambda_\textsc{QG}$ in the case of weakly-coupled string theory.
    
    We denote 4-cycles in $H_4\brap{X,\bbZ}$ by charges $\tilde q$ with tildes, and 2-cycles in $H_2\brap{X,\bbZ}$ by charges $q$ without tildes.
    We define a norm on these charges using $\metric$ and $\metricInvNoIndex$ respectively, namely
    \begin{equation}
        \bravv{\tildeq}:=\sqrt{\tildeq^i\metric_{ij}\tildeq^j}
        \,,~~~~
        \bravv{q}:=\sqrt{q_i\metricInvIndexed^{ij}q_j}
        \,.
    \end{equation}

\subsubsection{\texorpdfstring{$C_4$}{C4} axions}
    The kinetic term for $C_4$ axions $\theta_i$ is given by \cite{Grimm:2007hs}:
 	\begin{equation}
 		-\int \frac{\Mpl^2}{\brap{2\pi}^2}e^{2\phi}{\tilde G}^{i \bar j}\partial_\mu \theta_i \partial^\mu\theta_j\,,
 	\end{equation}
 	where $g_s=e^\phi$, and
 	\begin{align}
 		{\tilde G}^{i\bar j}
 		= \frac18\frac{t^it^j}{\cF^2} - \frac14\frac{\cF^{ij}}{\cF}
 		= \frac{1}{4\Vol}\metricInvIndexed^{ij}\,.
 	\end{align}
     Hence, the decay constant \eqref{eq:f-axion-string} for the axion with corresponding axion string obtained by wrapping a D3-brane on a 2-cycle $q \in H_2(X, \mathbb{Z})$ 
    is given by 
    \begin{equation}
        f_{P=q}^2
        =\frac{g_s^2\Mpl^2}{8\pi^2\Vol}
        \bravv{q}^2
        =\frac{M_s^2}{\brap{2\pi}^3}\bravv{q}^2
        \,.
        \label{fqC4}
    \end{equation}
    Note that we could have changed basis so that $q_i=\delta_i^j$ for some $j$ so as to more directly mirror \eqref{fdiag}, but this is in general not the case for the basis of curves that \CYTools{} naturally uses.
	
    A D3-brane instanton of charge ${\tilde q}^i$ has action given by
    \begin{equation}
        S = \frac{2\pi}{g_s} {\tilde q}^i\tau_i\,.
    \end{equation}
    
	Hence, from \eqref{rhoax},
    we have
    \begin{align}
        \rhoAx{4}{q,\tilde q}
        = \sqrt{
        g_s
            \frac{1}{
                \bravv{q}^2
                {{\tilde q}\cdot\tau}
            }
        }
        \,,
        \label{eqn.rho.C4.axion}
    \end{align}
    where we require $q$ and $\tildeq$ to have non-zero Dirac pairing.

    The generalization of the decay constant to multiple axions in \eqref{eq:f-instanton} for an instanton obtained by wrapping by wrapping a D3-brane on the divisor $\tildeq^iD_i$ is
    \begin{equation}
        f_{Q=\tildeq}^2
        =\frac{g_s^2\Mpl^2}{8\pi^2\Vol}
        \bravv{\tildeq}^{-2}
        =\frac{M_s^2}{\brap{2\pi}^3}\bravv{\tildeq}^{-2}
        \,,
        \label{fqtC4}
    \end{equation}
    so from \eqref{rhoinst}, we have
		\begin{align}
            \rhoInst{4}{\tilde q}
            = \sqrt{
                g_s
                \frac{
                    \bravv{\tildeq}^2
                }
                {
                    {{\tilde q} \cdot \tau}
                }
            }
            \,.
            \label{eqn.rho.C4.instanton}
		\end{align}
        
        We are interested in computing the maximum value of $\rhoAx{4}{q,\tilde q}$ and $\rhoInst{4}{\tilde q}$. From above, we see that these are maximized when $g_s$ is maximized. Since perturbative control requires $g_s <1$, we conservatively set $g_s=1$ throughout our analysis.

\subsubsection{\texorpdfstring{$C_2$}{C2} axions}
    The kinetic term for $C_2$ axions $c^i$ is given by \cite{Grimm:2007hs}:
 	\begin{equation}
 		-\int \frac{\Mpl^2}{\brap{2\pi}^2}e^{2\phi}{G}_{i \bar j}\partial_\mu c^i \partial^\mu c^j\,,
 	\end{equation}
 	where 
 	\begin{align}
 		G_{i \bar j}
 		= \frac14 \frac{\cF_i\cF_j}{\cF^2} - \frac14 \frac{\cF_{ij}}{\cF}
 		= {\frac1{4\Vol}} \metric_{ij}\,.
 	\end{align}
	From \eqref{fdiag}, we then have
    \begin{equation}
        f_{P=\tilde q}^2
        = \frac{g_s^2\Mpl^2}{8\pi^2\Vol}
        \bravv{\tildeq}^2
        =\frac{M_s^2}{\brap{2\pi}^3}\bravv{\tildeq}^2
        \,,
        \label{fqtC2}
    \end{equation}
    for the axion with corresponding axion string obtained by wrapping a D5-brane on the divisor ${\tilde q}^iD_i$.
    A D1-brane instanton of charge $q_i$ under these axions has instanton action
    \begin{equation}
        S = \frac{2\pi}{g_s}q_it^i\,.
    \end{equation}
	Hence, we find
    \begin{align}
        \rhoAx{2}{\tilde q,q}
        = \sqrt{
            g_s
            \frac{1 }{
            \bravv{\tildeq}^2
            q\cdot t
            }
        }
        \,,
        \label{eqn.rho.C2.axion}
    \end{align}
    where, as in the $C_4$ case, we shall require $q$ and $\tildeq$ to have non-zero Dirac pairing.

    The generalization of the decay constant to multiple axions in \eqref{eq:f-instanton} for an instanton obtained by wrapping by wrapping a D1-brane on the 2-cycle $q$ is
    \begin{equation}
        f_{Q=q}^2
        =\frac{g_s^2\Mpl^2}{8\pi^2\Vol}
        \bravv{q}^{-2}
        =\frac{M_s^2}{\brap{2\pi}^3}\bravv{q}^{-2}
        \,,
        \label{fqC2}
    \end{equation}
    so from \eqref{rhoinst}, we find
    \begin{align}
        \rhoInst{2}{q}
        = \sqrt{
            g_s
            \frac{
                \bravv{q}^2
            }{
                q \cdot t
            }
        }
        \,.
        \label{eqn.rho.C2.instanton}
    \end{align}
    
    As in the case of the $C_4$ axions, $\rhoAx{2}{\tilde q,q}$ and $\rhoInst{2}{q}$ are maximized when $g_s$ is maximized, so we may conservatively take $g_s=1$.

    \subsubsection{\texorpdfstring{$B_2$}{B2} axions}

    The case of $B_2$ axions is nearly identical to that of $C_2$ axions. We have
    \begin{align}
        f_{B_2} = \frac{1}{g_s}f_{C_2} ~\,~~~ S_{\text{F1}} = g_s S_{\text{D1}} \,,
    \end{align}
    where $S_{\text{F1}}$ is the action of a worldsheet instanton and $S_{\text{D1}}$ is the action of a D1-brane instanton wrapped around the same 2-cycle. Thus, we have
    \begin{align}
        \rho_{B_2} = \sqrt{g_s}\, \rho_{C_2}\,.
    \end{align}
        For $g_s < 1$, $\rho_{B_2}$ is always smaller than $\rho_{C_2}$, and for a given value of $g_s$ the two contain redundant information. Therefore, we ignore $B_2$ axions in what follows, focusing exclusively on $C_2$ axions and $C_4$ axions.

\section{Analytic results near facets}\label{sec.analytics}

    In this section, we give analytical arguments that $\rho$ remains order-one throughout the regions of type IIB moduli space under theoretical control.
    
	We are particularly interested in the behavior of the quantity $\rho$ near the boundaries of the K\"ahler cone, where the quantities of interest (instanton actions, tensions, decay constants, etc.)\ may diverge or vanish, and one could (in principle) imagine that $\rho$ might diverge. However, in order to maintain perturbative control, we must also insist 
    that all curve volumes remain large in string units, $t^i > 1$.

To simultaneously satisfy these two (seemingly contradictory) constraints, we restrict our attention to paths in moduli space of the form:
	\begin{equation}
		t^i = L\bras{ \tilde t^i + s\hat t^i}\,,
	\label{eqn:path}
    \end{equation}
	where $L \gg 1$, $L=s^{-1}$, and the point on the boundary is $t= \tilde t$. (Note that $L$ and $s$ are both dimensionless quantities.) This path represents a straight line in K\"ahler moduli space combined with an overall rescaling by $L = s^{-1}$, which ensures that the smallest curve volume remains order-one in the limit $s \rightarrow 0$, while the largest curve volumes diverge. Thus, while all curves remain sufficiently large in string units to maintain computational control, the relative sizes of different curves diverges along the path of interest. 

    As explained in \cite{Witten:1996qb} (see also \cite{Lee:2019wij, Alim:2021vhs, Brodie:2021ain, Gendler:2022ztv, BPSstrings}), boundaries of the K\"ahler cone can be classified into the following types:
\begin{enumerate}
    \item A curve collapses to a point.
    \item A divisor collapses to (a) a curve or (b) a point.
    \item The entire Calabi-Yau collapses.
    \item An infinite sequence of automorphisms occurs.
\end{enumerate}
These boundaries can be distinguished in part by their scaling behavior under a path of the form \eqref{eqn:path}. 
For boundaries of type 1, type 2(a), and type 2(b), the overall Calabi-Yau volume scales as $\mathcal{V} \sim L^3 s^0$. In contrast, boundaries of type 3 can be further sorted into two subcategories: either (a) $\mathcal{V} \sim L^3 s^1 \sim L^2$ or (b) $\mathcal{V} \sim L^3 s^2 \sim L^1$.\footnote{In M-theory compactifications, these subcategories of type 3 boundaries correspond to decompactification limits and emergent string limits, respectively.} We may safely ignore boundaries of type 4, as any path approaching such a boundary traverses the fundamental domain of moduli space infinitely many times.\footnote{Said differently, type 4 boundaries, also known as ``periodic boundaries,'' represent boundaries of the marked moduli space \cite{Raman:2024fcv} that are not boundaries of the (fundamental domain of) moduli space.} This means that any point near such a boundary is gauge-equivalent to a point that is far from that boundary.

In principle, parametric violations of the string scale bound \eqref{stringscalebound} might arise in limits with vanishing $f/M_s$ or vanishing $S$. However, along paths of the form \eqref{eqn:path}, instanton actions always remain order-one or larger, so potential violations come solely from vanishing axion decay constants, $f/M_s \rightarrow 0$.
The sources of potential violations at the various types of boundaries are shown in Table \ref{tab:potentialviolations}. 

\begin{table}
    \centering
    \begin{tabular}{c|c|c|c|c} 
    \multirow{2}{*}{Boundary type} &   \multicolumn{2}{c|}{$C_4$ axions} & \multicolumn{2}{c}{$C_2$ axions} \\ \cline{2-5}   &  
    $f \rightarrow 0$? &  $S \rightarrow 0$?  &   $f \rightarrow 0$?   &  $S \rightarrow 0$? \\ \hline 
     Type 1 & no & no &  no & no \\
     Type 2a & no &  no & no  & no \\
     Type 2b & yes  & no & no  & no \\
     Type 3a & yes  & no & no & no \\
     Type 3b & yes  & no & yes & no  \\
     Type 4 & n/a & n/a & n/a & n/a \\
    \end{tabular}
    \caption{Sources of potential violations of the string scale bound \eqref{stringscalebound} near various boundaries of the K\"ahler cone due to vanishing $f$ or vanishing $S$ (measured in string units, $M_s = 1$).
    Periodic boundaries are boundaries of the marked moduli space \cite{Raman:2024fcv} but not the moduli space, so they can be ignored by a suitable gauge transformation. Note that instanton actions never vanish along paths of the form \eqref{eqn:path}. }
  \label{tab:potentialviolations}
\end{table}

From \eqref{fqC4} and \eqref{fqtC4}, we have that $C_4$ axion decay constants, as measured in string units, are related to $\bravv{q}$ and $\bravv{\tilde q}$ via
\begin{equation}
\frac{f_{P=q}}{M_s} \sim \bravv{q}\,,~~~\frac{f_{Q=\tilde q}}{M_s} \sim \bravv{\tilde q}^{-1}\,.
\end{equation}
Correspondingly, a divergence of $\rhoInst{4}{\tilde q}$ cannot occur unless $\bravv{\tilde q} \sim M_s / f_{Q=\tilde q}$ vanishes, and a divergence of $\rhoAx{4}{q, \tilde q}$ cannot occur unless $\bravv{q} \sim f_{P=q}/M_s$ diverges.

Similarly, from \eqref{fqC2} and \eqref{fqtC2}, we see that $C_2$ axion decay constants are related to $\bravv{q}$ and $\bravv{\tilde q}$ via
\begin{equation}
\frac{f_{Q=q}}{M_s} \sim \bravv{q}^{-1}\,,~~~\frac{f_{P=\tilde q}}{M_s} \sim \bravv{\tilde q}\,.
\end{equation}
Correspondingly, a divergence of $\rhoInst{2}{q}$ cannot occur unless $\bravv{q} \sim M_s / f_{Q=q}$ vanishes, and a divergence of $\rhoAx{2}{\tilde q, q}$ cannot occur unless $\bravv{\tilde q} \sim f_{P=\tilde q}/M_s$ diverges.

While the vanishing or diverging of $\bravv{q}$ or $\bravv{\tilde q}$ is a necessary condition to parametrically violate the string scale bound \eqref{stringscalebound}, it is not a sufficient condition. To investigate these potential violations further, following \cite{Reece:2025zva},\footnote{In \cite{Reece:2025zva}, the authors carried out an analogous computation within the K\"ahler moduli space of a Calabi-Yau compactification of M-theory. After accounting for the overall scaling factor $L = s^{-1}$, these results translate directly to scaling behavior of instantons and instanton strings in the Type IIB case at hand. Related results on scaling behavior also appeared recently in~\cite{Blanco:2025qom} (see also~\cite{Marchesano:2023thx,Marchesano:2024tod,Castellano:2024gwi}).} we sort BPS instantons and BPS axion strings based on their scaling behaviors at each type of boundary. These scaling behaviors are determined by the subspace in which their charges reside. This, in turn, determines the scaling behavior of the quantity of interest, $\rho$.

    More precisely, let us define the following subspaces of the instanton/axion string charge lattices:
    \begin{align}
        \cMlight &:= \brac{q:q_a {\tilde t}^a=0}
        \,,&
        \cElimdir &:={\cMlight}^\orthogcomplement = \brac{{\tilde q}: {\tilde q}^i \propto {\tilde t}^i} \label{MLdef}
        \,,\\
        \cEperp &:= \brac{{\tilde q}: {\tilde t}^a{\tilde t}^bC_{abi}{\tilde q}^i=0}
        \,,&
        \cMtau &:= {\cEperp}^\orthogcomplement = \brac{q: q_i \propto {\tilde t}^a{\tilde t}^bC_{abi}}
        \,,\\
        \cEker &:= \brac{{\tilde q}: {\tilde q}^i \in \ker\brap{{\tilde t}^aC_{aij}}}
        \,,&
        \cMkerorth &:= {\cEker}^\orthogcomplement \label{Ekerdef}
        \,.
    \end{align}
   Along a path of the form \eqref{eqn:path} with $s= 1/L$, we will then observe the scaling behavior
    \begin{equation}
   \bravv{q}^2 \sim s^{\lambda_q}/L \,,~~~~ \bravv{\tilde q}^2 \sim s^{\lambda_{\tilde q}} L\,,~~~~ \rho^2 \sim L^{\alpha}\,,
    \end{equation}
where the scaling coefficients $\lambda_q$, $\lambda_{\tilde q}$, $\alpha$ depend on the subspace(s) in which the instanton charge resides. Potential violations of the string scale bound \eqref{stringscalebound} occur only when $\bravv{q}^2$ or $\bravv{\tilde q}^2$ vanish or diverge in the limit $L \rightarrow \infty$; actual violations occur only when $\alpha > 0$.

\subsection{Classification of scaling behavior}\label{ss.classification}
 
\begin{table}[ht]
    \centering
    \begin{tabular}{|c|cllrr|}
        \hline
        ${\Vol}$ & ${\tilde q}$ for this behavior & $\bravv{\tilde q}^2$ & $S_{\tildeq}$ & $\alphaInst{4}{\tilde q}$ & $\alphaAx{4}{q,\tilde q}$\\
        \hline
        \multirow{3}{*}{$L^3s^2$}
            &${\tilde q} \in \cElimdir$
                & $Ls^{2}$ & $L^2s^{2}$ & $-1$    & $0$\\
            &${\tilde q}\in\cEker$, but ${\tilde q} \notin \cElimdir$
                & $Ls^{1}$    & $L^2s^{2}$    & $0$    & $0$\\
            &${\tilde q}\notin\cEker$
                & $Ls^{0}$    & $L^2s^{1}$    & $0$    & $0$\\
        \hline
        \multirow{4}{*}{$L^3s^1$}
            &${\tilde q} \in \cEker$
                & $Ls^{1}$ & $L^2s^2$ & $0$    & $0$\\
            &${\tilde q} \in \cEker\oplus\cElimdir$, but ${\tilde q} \notin\cEker$
                & $Ls^{1}$ & $L^2s^1$ & $-1$    & $-1$\\
            &${\tilde q} \in \cEperp$, but ${\tilde q} \notin \cEker\oplus\cElimdir$
                & $Ls^{0}$ & $L^2s^1$ & $0$    & $0$\\
            &${\tilde q} \notin \cEperp$
                & $Ls^{-1}$ & $L^2s^0$ & $0$    & $0$\\
        \hline
        \multirow{3}{*}{$L^3s^0$}
            &${\tilde q}\notin \cEperp$
                & $Ls^{0}$           & $L^2s^{0}$       & $-1$    & $-1$\\
            &${\tilde q}\in \cEperp$, but ${\tilde q} \notin \cEker$
                & $Ls^{0}$           & $L^2s^{1}$       & $0$    & $0$\\
            &${\tilde q} \in \cEker$
                & $Ls^{1}$     & $L^2s^{2}$   & $0$    & $0$\\
        \hline
    \end{tabular}
    \caption{
        The possible scaling behaviors of $
        \bravv{\tilde q}^2$, the action $S_{\tilde q}$ of an instanton of charge $\tildeq$ under $C_4$ axions, $\brap{\rhoInst{4}{\tilde q}}^2 \sim L^{\alphaInst{4}{\tilde q}}$, and $\brap{\rhoAx{4}{q,\tilde q}}^2 \sim L^{\alphaAx{4}{q,\tilde q}}$ for a given scaling behavior of $\Vol$.
        For $\alphaAx{4}{q,\tilde q}$, the table only lists the largest such $\alpha$ across all $q$ of non-zero Dirac-pairing with the given $\tilde q$, where the $q$ are taken from \autoref{table.scaling.C2} (restricting to the section with the same parametric scaling of $\Vol$).
    }
	\label{table.scaling.C4}
\end{table}
\begin{table}[ht]
    \centering
    \begin{tabular}{|c|cllrr|}
        \hline
        ${\Vol}$ & ${q}$ for this behavior & $\bravv{q}^2$ & $S_{q}$ & $\alphaInst{2}{q}$ &  $\alphaAx{2}{\tilde q, q}$\\
        \hline
        \multirow{3}{*}{$L^3s^2$}
            &$q \notin \cMlight$
                & $\frac1L s^{-2}$    & $L s^{0}$    & $0$    & $0$\\
            &$q \in \cMlight$, but $q \notin \cMkerorth$
                & $\frac1L s^{-1}$    & $L s^{1}$    & $0$    & $0$\\
            &$q\in \cMkerorth$ (so also $q \in \cMlight$)
                & $\frac1L s^{0}$    & $L s^{1}$    & $-1$    & $-1$\\
        \hline
        \multirow{4}{*}{$L^3s^1$}
            &$q \notin \cMkerorth \cap \cMlight$ but $q \in \cMlight$
                & $\frac1L s^{-1}$ & $L s^{1}$ & $0$    & $0$\\
            &$q \notin \cMlight$
                & $\frac1L s^{-1}$ & $L s^{0}$ & $-1$    & $-1$\\
            &$q \in \cMkerorth \cap \cMlight$, but $q \notin \cMtau$
                & $\frac1L s^{0}$ & $L s^{1}$ & $-1$    & $-1$\\
            &$q \in \cMtau$
                & $\frac1L s^{1}$ & $L s^{1}$ & $-2$    & $-2$\\
        \hline
        \multirow{4}{*}{$L^3s^0$}
            &$q\in \cMkerorth$ and $q \notin \cMlight$
                & $\frac1L s^{0}$           & $L s^{0}$       & $-2$    & $-2$\\
            &$q\in \cMkerorth$ and $q \in \cMlight$
                & $\frac1L s^{0}$           & $L s^{1}$       & $-1$    & $-1$\\
            &$q\notin \cMkerorth$ and $q \in \cMlight$
                & $\frac1L s^{-1}$    & $L s^{1}$       & $0$    & $0$\\
            &$q\notin \cMkerorth$ and $q \notin \cMlight$
                & $\frac1L s^{-1}$    & $L s^{0}$       & $-1$    & $-1$\\
        \hline
    \end{tabular}
    \caption{
        The possible scaling behaviors of $\bravv{q}^2$, the action $S_{q}$ of an instanton of charge $q$ under $C_2$ axions, $\brap{\rhoInst{2}{q}}^2 \sim L^{\alphaInst{2}{q}}$, and $\brap{\rhoAx{2}{\tilde q, q}}^2 \sim L^{\alphaAx{2}{\tilde q, q}}$ for a given scaling behavior of $\Vol$.
        For $\alphaAx{2}{\tilde q, q}$, the table only lists the largest such $\alpha$ across all $\tilde q$ of non-zero Dirac-pairing with the given $q$, where the $\tilde q$ are taken from \autoref{table.scaling.C4} (restricting to the section with the same parametric scaling of $\Vol$).
    }
    \label{table.scaling.C2}
\end{table}

In \autoref{table.scaling.C4} and \autoref{table.scaling.C2}, we display all possible scaling coefficients $\alpha$ for 
D3-brane instantons and D1-brane instantons, respectively. The crucial observation is that all BPS instantons satisfy
\begin{equation}
\alpha \leq 0\,.
\end{equation}
This implies that in the limit $L \rightarrow \infty$, $\rhoInst{4}{q}$, $\rhoAx{4}{q,\tilde q}$, $\rhoInst{2}{q}$, and $\rhoAx{2}{\tilde q,q}$ all remain finite.

Recall that the physically observable axion decay constant $f$ is not, in general, given simply by the decay constants $f_{Q} = 1/\sqrt{Q_i\kappa^{ij}Q_j}$ or $f_{P} = \sqrt{{P}^i\kappa_{ij}{P}^j}$ studied here (and defined above in \S\ref{ss.proxies}). In principle, then, one could imagine that the observable decay constant $f$ be parametrically suppressed relative to $f_{Q}$, $f_{P}$ for any rational $Q, P$.
In the Calabi-Yau context at hand, however, this issue seems to be irrelevant.
To see this, we first recall from \S\ref{ss.challenge}-\ref{ss.proxies} that a parametrically suppressed effective decay constant $f_{\rm eff} \ll M_s$ requires a small $f_{P} \ll M_s$. By \eqref{fqtC2} and \eqref{fqC4}, this requires $\bravv{\tilde q} \ll 1$ (for $C_2$ axions) or $\bravv{q} \ll 1$ (for $C_4$ axions). Per \autoref{table.scaling.C4} and \autoref{table.scaling.C2}, this scaling behavior occurs only for the following cases:
\begin{align}
\cV \sim L^3 s^2&:~~~~\tilde q \in \cE_0 ~~\text{ or } ~~q \in \cMkerorth   \label{eq:L3s2}\\
\cV \sim L^3 s^1&:~~~~ q \in \cMkerorth  \cap \cMlight \setminus \cMtau~~\text{ or } ~~ q \in \cMtau  \\
 \cV \sim L^3 s^0 &:~~~~ q \in \cMkerorth \,.
 \label{eq:L3s0}
\end{align}

From here, we may invoke a relevant bit of mathematical lore: the \emph{cone conjectures} of Morrison \cite{Morrison93, Morrison94} and the dual coordinate cone conjecture \cite{BPSstrings} ensure that all of the subspaces appearing in \eqref{eq:L3s2}-\eqref{eq:L3s0} are rational polyhedral cones,%
\footnote{
    In particular, the cone conjectures ensure that
    $\cE_0$ and $\cEker$ are rational polyhedral in limits with $\Vol \sim L^3s^2$,
    $\cE_0 \oplus \cEker$ and $\cMtau$ are rational polyhedral in  limits with $\Vol \sim L^3s^1$,
    and $\cEker$ is rational polyhedral in limits with $\Vol \sim L^3s^0$.
    Since $\cMkerorth = \cEker^\orthogcomplement$ and $(\cMkerorth \cap \cMlight) = (\cE_0 \oplus \cEker)^\orthogcomplement$, the former are rational polyhedral whenever the latter are.%
} %
i.e., they are generated by a finite number of integral charges \cite{BPSstrings}.\footnote{In 5d M-theory compactifications, these charges $\tilde q^i$, $q_i$ correspond to charges of strings and particles, respectively, that become light at the boundary of the K\"ahler cone. These strings and particles are needed to satisfy the Emergent String Conjecture \cite{Lee:2019wij}, the Distance Conjecture \cite{Ooguri:2006in}, and the Weak Gravity Conjecture \cite{Arkanihamed:2006dz}.}

 The upshot of this is that the subspaces of charges $q$, $\tilde q$ associated with vanishing decay constants $f_{P} /M_s \rightarrow 0$ are generated by integral charge vectors, hence they fall under the purview of our study. If the QCD axion is a linear combination of Type IIB axions, we expect that it will obey the bound \eqref{stringscalebound}.

\subsection{Physical interpretation of scaling behavior}\label{ss.interpretation}

	\begin{table}
		\centering
        \def\arraystretch{1.3}
        \begin{tabular}{|c|cc|cc|}
            \hline
           \multirow{2}{*}{Boundary type} & \multicolumn{4}{c|}{Maximum value of $\alpha$}\\ \cline{2-5}
            & $\alphaInst{4}{\tilde q}$ & $\alphaAx{4}{q,\tilde q}$ & $\alphaInst{2}{q}$ & $\alphaAx{2}{\tilde q,q}$ \\\hline
            1 & $-1$ & $-1$ & $-1$ & $-1$ \\ \hline
            2a & 0 & 0 & $-1$ & $-1$ \\\hline
            2b & 0 & 0 & 0 & 0 \\\hline
            3 & 0 & 0 & 0 or $-1$ & 0 or $-1$ \\\hline
        \end{tabular}
		\caption{Maximum scaling coefficients $\alpha$ for paths approaching different types of boundaries of the K\"ahler cone.}
	\label{table.absent.or.present}
	\end{table}

Further information about scaling of the largest value of $\rho$ in a given limit can be deduced based on the type of boundary in question. In particular, one can argue that the subsets of $H_4(X, \mathbb{R})$ with $\alphaBoth{4} = 0$ are nonempty in type 3 limits (i.e., limits with $\cV \sim L^3 s^1$ or $\cV \sim L^3 s^2$) \cite{Reece:2025zva}. Furthermore, the Emergent String Conjecture \cite{Lee:2019wij} requires the existence of effective divisors with $\tilde q \notin \cEker$,\footnote{In an M-theory compactification to 5d, these effective divisors correspond to KK monopoles in some duality frame.} which have $\alphaBoth{4} = 0$. This suggests that every type 3 limit will feature D3-brane instantons with $\alphaInst{4}{\tilde q} =\alphaAx{4}{q,\tilde q} = 0$.

Similarly, the divisors that collapse in boundaries of type 2(a)-(b) also have $\alpha = 0$. (More precisely, a divisor that collapses to a curve has $\tilde q \in \cEperp\setminus \cEker$, while a divisor that collapses to a point has $\tilde q \in \cEker$.) Thus, every type 2 limit features BPS D3-brane instantons with $\alphaInst{4}{\tilde q} = \alphaAx{4}{q,\tilde q} = 0$. In contrast, no effective divisors collapse at flop transition boundaries of type 1, so all BPS D3-brane instantons have $\alphaBoth{4} =-1$.

The analogous story for D1-brane instantons is a bit more complicated. Type 3 boundaries with $\cV \sim L^3 s^2$ always feature D1-brane instantons with $\alphaBoth{2} = 0$,\footnote{In M-theory compactifications to 5d, M2-branes wrapping these curves give rise to particles that may, in an appropriate duality frame, be interpreted as wrapped NS5-branes.} but in limits with $\cV \sim L^3 s^1$, the maximum value of $\alphaBoth{2}$ may be either $-1$ or $0$.\footnote{In M-theory compactifications to 5d, the case $\alphaBoth{2}=0$ occurs if the decompactification limit $s\rightarrow 0$ features a decoupled 6d SCFT; in contrast, a decompactification limit without a decoupled 6d SCFT has only $\alphaBoth{2}=-1$.}

Similarly, type 2(b) boundaries necessarily feature D1-brane instantons with $\alphaBoth{2} = 0$.\footnote{In M-theory compactifications to 5d, these curves correspond to electrically charged particles that become light at the boundary of interest, where a 5d SCFT decouples. Such particles may be interpreted as oscillation modes of a tensionless SCFT string \cite{Reece:2025zva}.} In contrast, in limits approaching type 1 and type 2(a) boundaries, there exist D1-brane instantons with $\alphaBoth{2} = -1$ but none with $\alphaBoth{2} = 0$.

These results are summarized in \autoref{table.absent.or.present}.

The fact that $\rhoInst{2}{q}$, $\rhoAx{2}{\tilde q,q}$ remain finite in limits approaching type I (flop transition) boundaries is especially noteworthy. In \cite{Reece:2025zva}, the authors noted that the co-scaling relation~\eqref{eq:axioncoscaling} is violated at such boundaries because a curve shrinks to zero size while all effective divisors remain large. In the type IIB case considered here, this means that an instanton action becomes small (i.e., $S \ll L$) without a corresponding axion string of tension $T \ll L^2 M_s^2$.
This shows that co-scaling is not necessary to ensure that $\rho$ remains finite.

\section{Numerical results}
\label{sec:numerics}

	In this section, we present numerical results from a survey of Calabi-Yau manifolds using \CYTools{}.

    Our survey proceeds as follows: first, for a given choice of $h^{1,1}$, we select a reflexive polytope in the Kreuzer-Skarke dataset~\cite{Kreuzer:2000xy}; this corresponds to a choice of Calabi-Yau geometry. Next, we randomly select a triangulation of the polytope; this corresponds to a choice of phase of the geometry in question, i.e., a choice of the K\"ahler cone.

    \begin{figure}[ht]
		\centering
		\begin{subfigure}{0.49\textwidth}
			\centering
			\includegraphics[width=\textwidth]{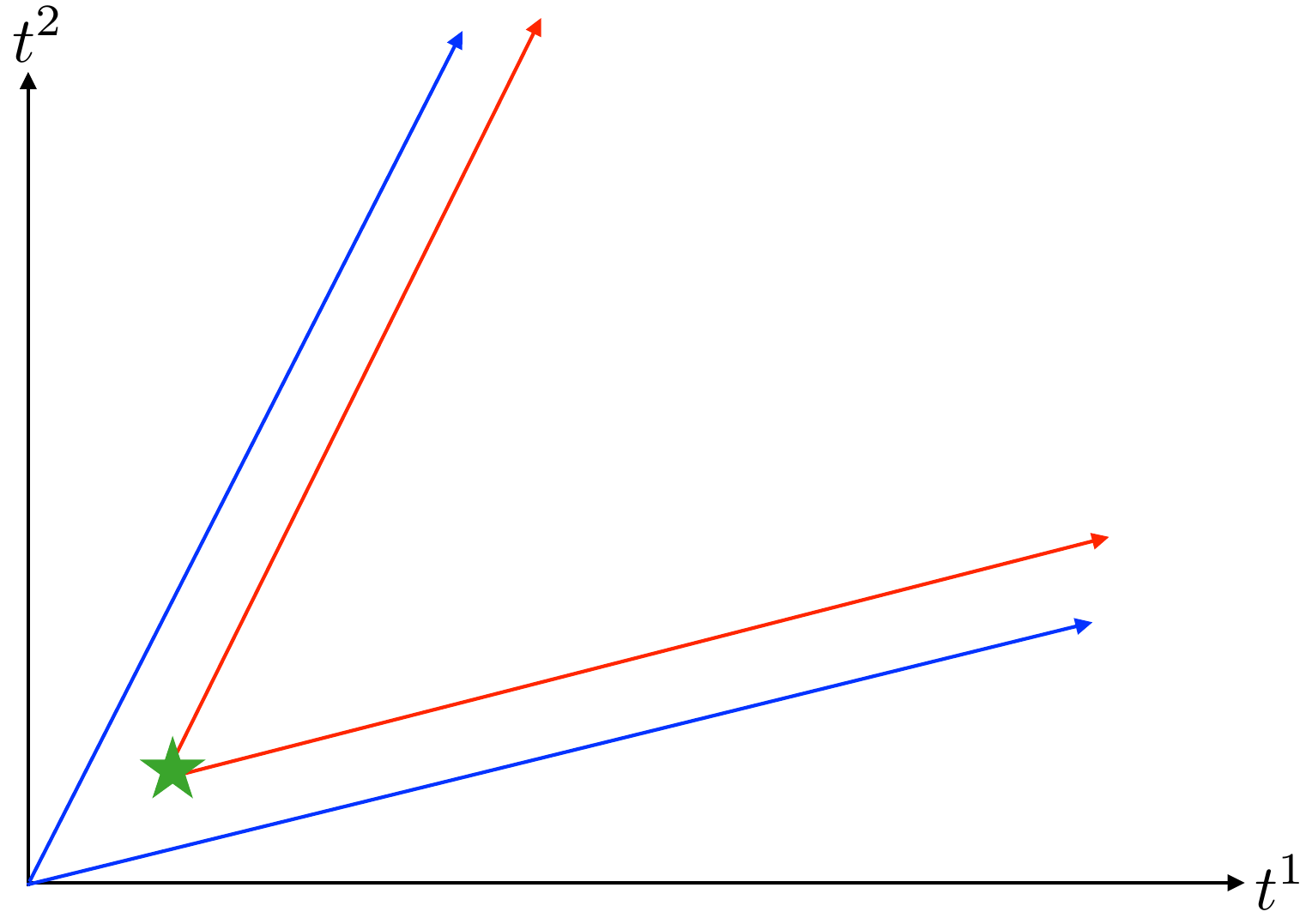}
			\caption{Tip of the stretched K\"ahler cone.}
            \label{sfig.stretched}
		\end{subfigure}
		\centering
		\begin{subfigure}{0.49\textwidth}
			\centering
	\includegraphics[width=\textwidth]{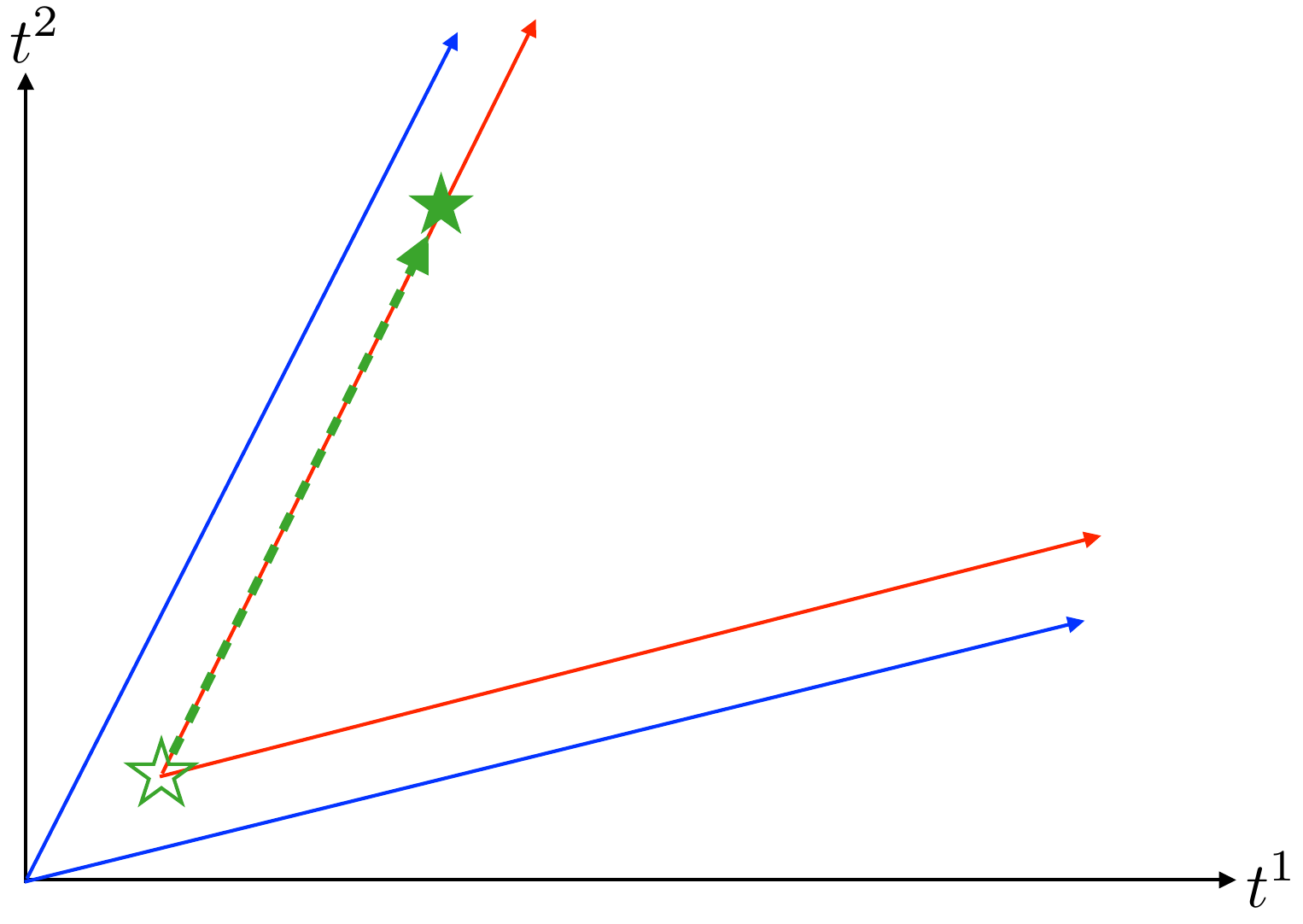}
			\caption{Near a facet.}
            \label{sfig.nearfacet}
		\end{subfigure}
		\caption{Points of the K\"ahler cone (blue) considered in our survey. (a) The tip of the stretched K\"ahler cone (red) is the point closest to the origin at which all curves have string-scale volume greater than 1. (b) Moving along a facet of the stretched K\"ahler cone, we find a point that is parametrically closer to that facet than to another facet.}
		\label{figure.stretched}
	\end{figure}

    We then calculate $\rho^2$ at (a) \CYTools{}'s approximation of the tip of the stretched K\"ahler cone and (b) in the neighborhood of a given facet. Here, the stretched K\"ahler cone is defined as the point closest to the origin such that all facets are at least a distance $1$ away (in string units). The tip of the stretched K\"ahler cone is depicted visually in \autoref{sfig.stretched}.
    
    To obtain a point in the neighborhood of a given facet, we begin at the tip of the stretched K\"ahler cone and move along a facet, ultimately arriving at a point that is parametrically closer to that facet than to another facet.%
    \footnote{
        More precisely, our algorithm began at the point $\vec s =$ \texttt{tip\tttunderscore{}of\tttunderscore{}stretched\tttunderscore{}cone}, obtained via \CYTools{}.
        If $\vec s$ was too far from or close to one of the facets, then the entire CY geometry was skipped over.
        Then, for a given reference facet with unit normal $\vec n$, it attempted to move away from the tip in the direction $\vec l= \vec s-\brap{\vec s \cdot \vec n}\vec n$, increasing the distance from other facets while maintaining the distance from the reference facet, until the furthest facet was a distance $L=10^{4}$ away, and the closest facet was a distance between 1 and 10 away.
        If moving in $\vec l$ in fact decreased the distance from one of the facets, or no distance satisfied the constraints on the minimum and maximum facet distances, then the facet was simply skipped.
    }
    We repeat this process for multiple facets of the stretched K\"ahler cone. This process is shown in \autoref{sfig.nearfacet}.

At each point in moduli space, we then compute the maximum value of $\rho^2$ for the four choices of $\rho$ given in \eqref{eqn.rho.C4.instanton}, \eqref{eqn.rho.C4.axion}, \eqref{eqn.rho.C2.instanton}, and \eqref{eqn.rho.C2.axion}, respectively. More precisely, we have
    \begin{align}
        \brap{\rhoInst{4}{\max}}^2 &:=
        \max_{\tilde q\in G_4}\brac{
            {\frac{ \bravv{\tildeq}^2 }{\tildeq \cdot \tau}}
        }
        \\
        \brap{\rhoAx{4}{\max}}^2 &:=
        \max_{q\in G_2}\max_{\tilde q\in G_4 : {q}_i{\tilde q}^i\neq0}\brac{
            \frac{1}{\bravv{q}^2 \tildeq \cdot \tau}
        }
        \,,
    \end{align}
and
    \begin{align}
        \brap{\rhoInst{2}{\max}}^2&:=
        \max_{q\in G_2}\brac{
            \frac{ \bravv{q}^2 }{ q \cdot t}
        } \\
        \brap{\rhoAx{2}{\max}}^2 &:=
        \max_{\tilde q \in G_4}\max_{q\in G_2:{\tilde q}^iq_i \neq 0}\brac{
            \frac{1}{\bravv{\tildeq}^2 q \cdot t}
        }
        \,.
    \end{align}
    Here, we have conservatively set $g_s = 1$, as discussed above.

    In the above formulae, if \texttt{cy} is the \texttt{CalabiYau} object representing the Calabi-Yau in \CYTools{}, then $G_2$ and $G_4$ are the sets of charges given by:
    \begin{align}
        G_2 &= \texttt{cy.\allowbreak{}toric\tttunderscore{}mori\tttunderscore{}cone(in\tttunderscore{}basis=True).\allowbreak{}rays()}
        \\
        G_4 &= \texttt{cy.\allowbreak{}toric\tttunderscore{}effective\tttunderscore{}cone().\allowbreak{}rays()}
    \end{align}
    The \CYTools{} functions \texttt{toric\tttunderscore{}mori\tttunderscore{}cone} and \texttt{toric\tttunderscore{}effective\tttunderscore{}cone} give approximations of the true Mori and effective cone respectively based on the corresponding cone of the ambient toric variety.
    The \texttt{rays} function then returns a set of charges that can be used to generate the cone it is applied to, but in general includes some redundant charges -- that is, a smaller set would still generate the cone. The \texttt{extremal\tttunderscore{}rays} function would produce the corresponding minimal generating set, but is computationally expensive at large cone dimension, so we do not use it.
    
    Note that, by restricting to $G_2$ and $G_4$, we do not test all possible charges.
    Indeed, there are in general infinitely many charges, so it would be infeasible to do so.
    However, we have examined 3586 facets in Calabi-Yau geometries with $h^{1,1} \in \brac{2,3,4,5}$ in detail and verified that in all these cases, the elements of $G_2$ and $G_4$ represent all possible scaling behaviors of $f$ near the boundary in question. In this sense, they offer a reasonably representative sample for the full charge lattice.

    \subsection{Tip of the stretched K\"ahler cone}

	\begin{figure}[!ht]
		\centering
		\begin{subfigure}{0.49\textwidth}
			\centering
			\includegraphics[width=\textwidth]{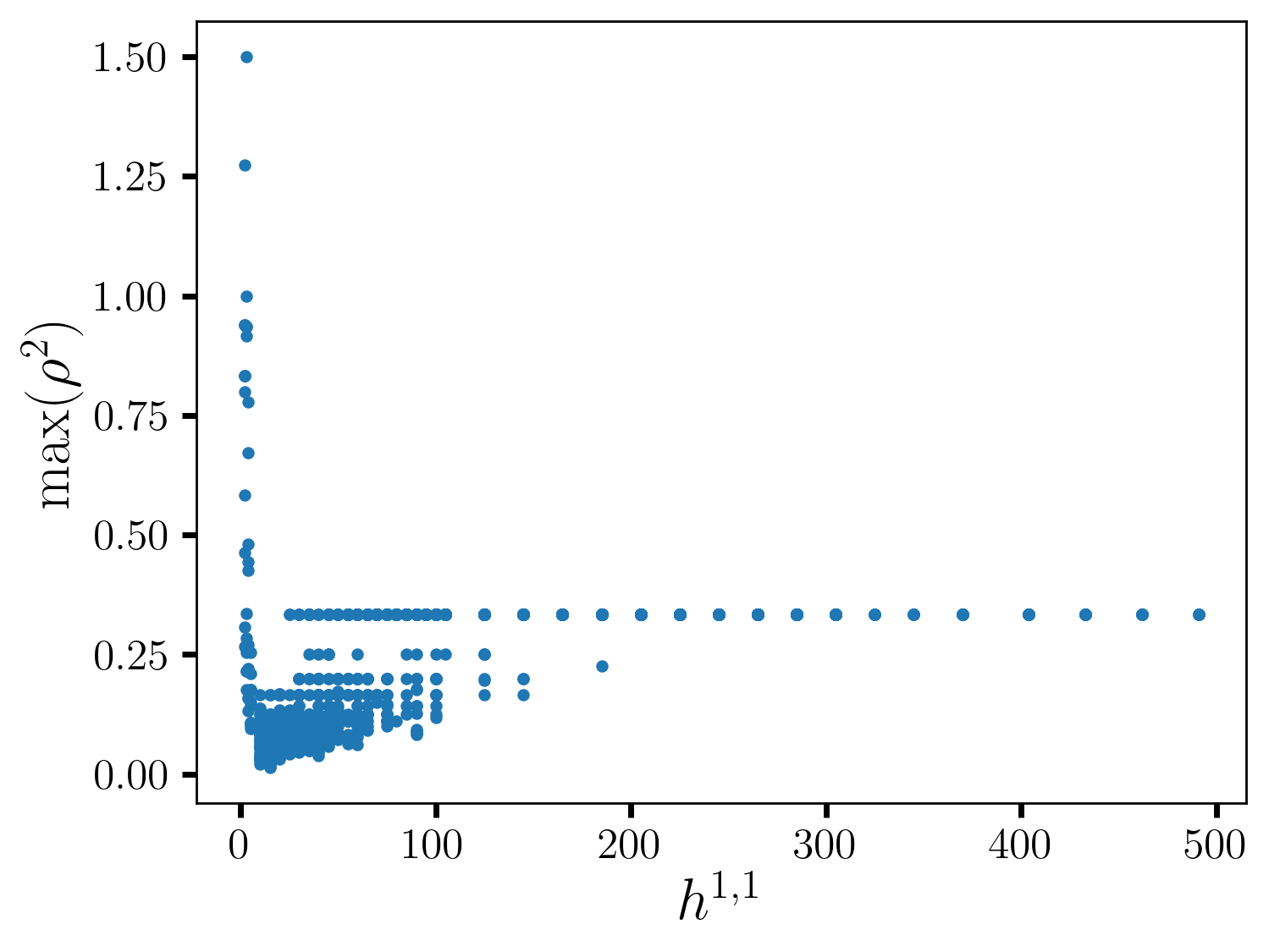}
			\caption{$C_2$; $f=f_{P=\tildeq}$.}
		\end{subfigure}
		\centering
		\begin{subfigure}{0.49\textwidth}
			\centering
			\includegraphics[width=\textwidth]{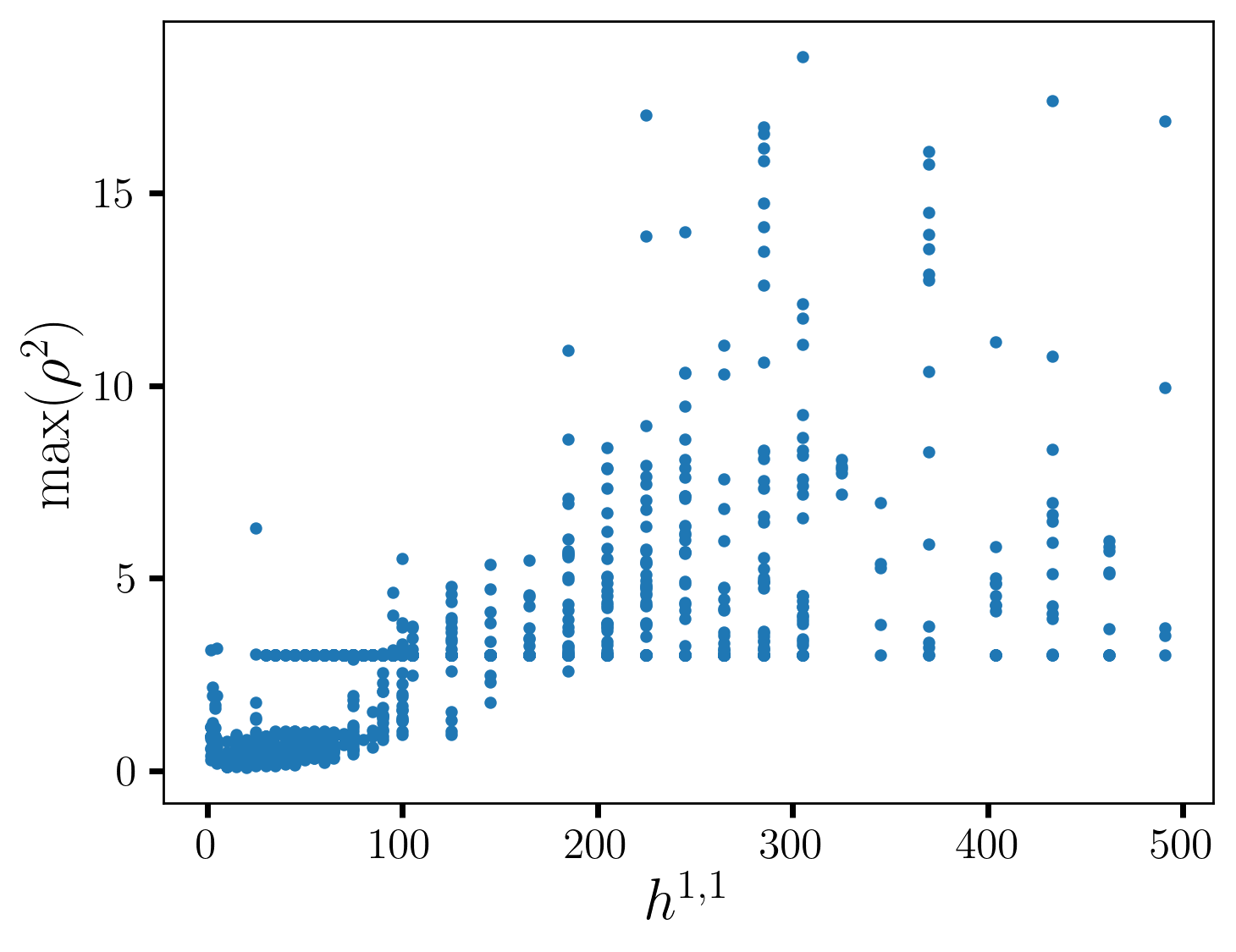}
			\caption{$C_2$; $f=f_{Q=q}$.}
		\end{subfigure}
		\begin{subfigure}{0.49\textwidth}
			\centering
			\includegraphics[width=\textwidth]{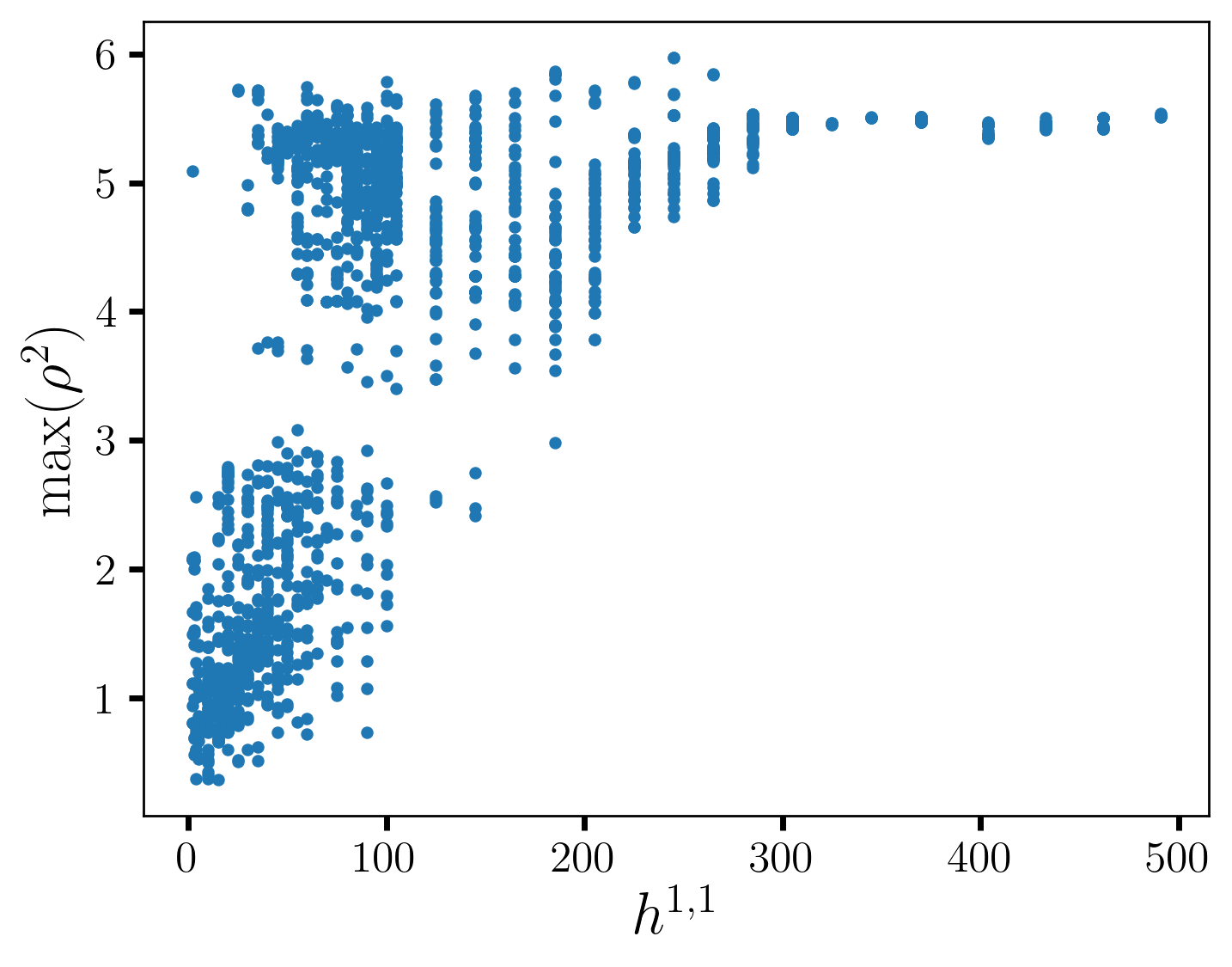}
			\caption{$C_4$; $f=f_{P=q}$.}
		\end{subfigure}
		\centering
		\begin{subfigure}{0.49\textwidth}
			\centering
			\includegraphics[width=\textwidth]{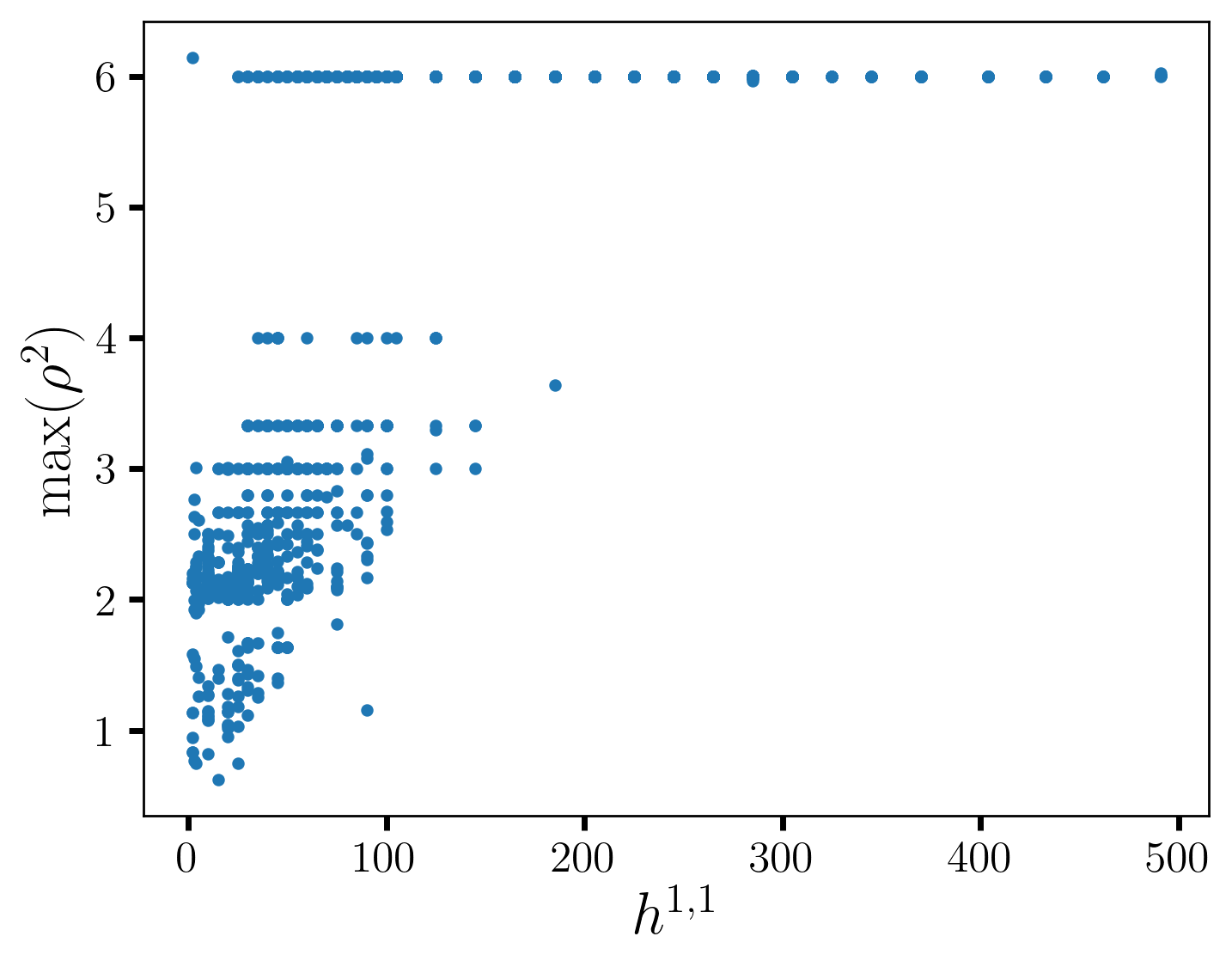}
			\caption{$C_4$; $f=f_{Q=\tildeq}$.}
		\end{subfigure}
		\caption{Plots of the maximum $\rho^2$, for different definitions of $\rho$, both $C_2$ and $C_4$ axions, where each point is at the tip of the stretched K\"ahler cone. Note the linear scale on the y-axis.}
		\label{plots.rho.tip.all.linear.axis}
	\end{figure}

	\autoref{plots.rho.tip.all.linear.axis} shows the plots of $\rho_{\rm max}$ at the tip of the stretched K\"ahler cone. In total, we scanned 1590 Calabi-Yau manifolds, with $h^{1,1}$ ranging from $2$ to $491$. The largest value of $\rho^2$ observed was $18.5$ (to 1 d.p.), corresponding to $\rhoInst{2}{q} \approx 4.3$.

    Patterns appear in the plots for large $h^{1,1}$.
	In particular, for large $h^{1,1}$, $(\rhoAx{2}{\max})^2$ and $(\rhoInst{4}{\max})^2$ seem to be approximately $\frac13$ and $6$ respectively, and $(\rhoInst{2}{\max})^2$ seems to be bounded below by approximately $3$. Partial explanations of these repeated values are explained below in Appendix \ref{sec.structure.of.tip.plots}.

\subsection{Near facets}

	\begin{figure}[!ht]
		\centering
		\begin{subfigure}{0.49\textwidth}
			\centering
			\includegraphics[width=\textwidth]{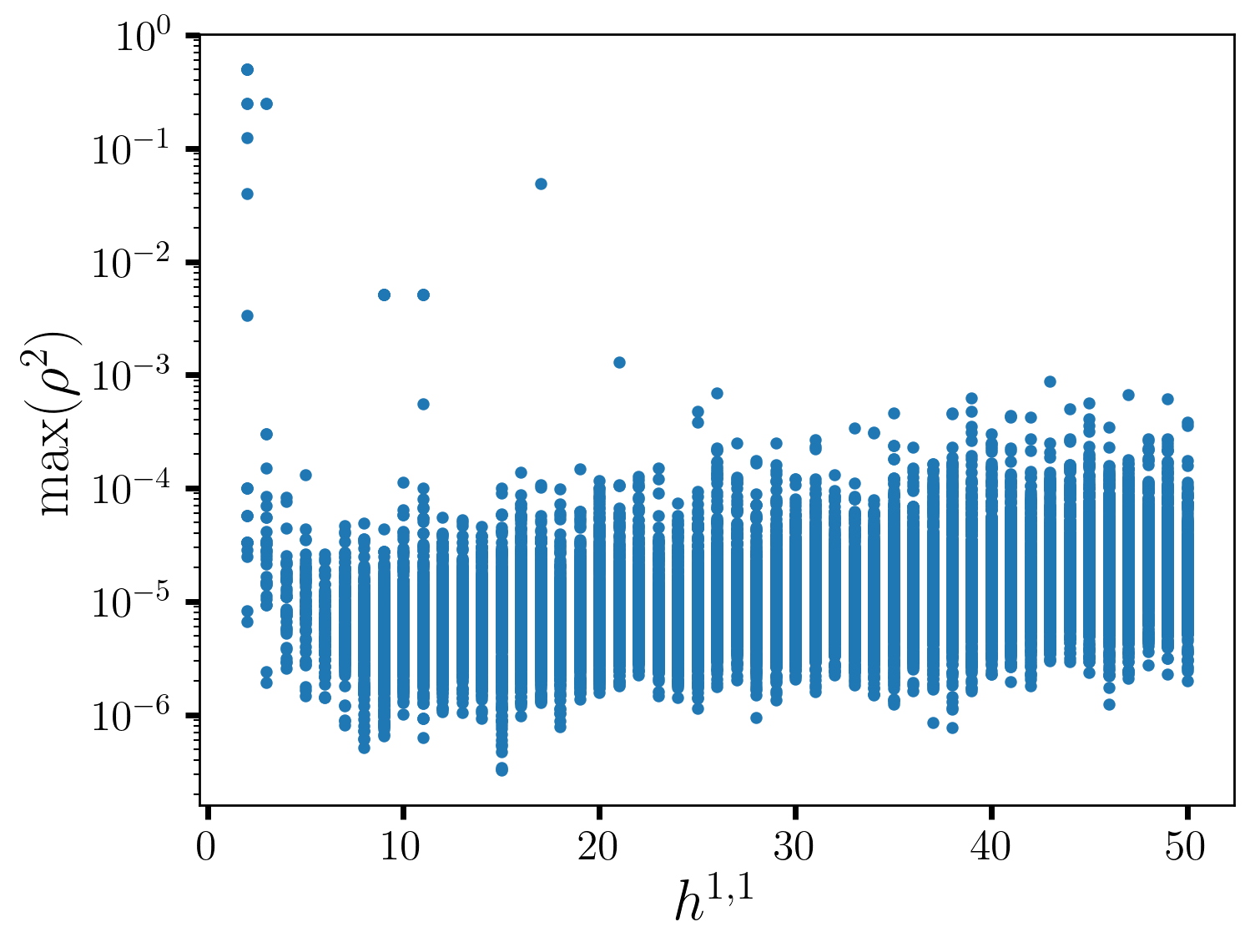}
			\caption{$C_2$; $f=f_{P=\tildeq}$.}
		\end{subfigure}
		\centering
		\begin{subfigure}{0.49\textwidth}
			\centering
			\includegraphics[width=\textwidth]{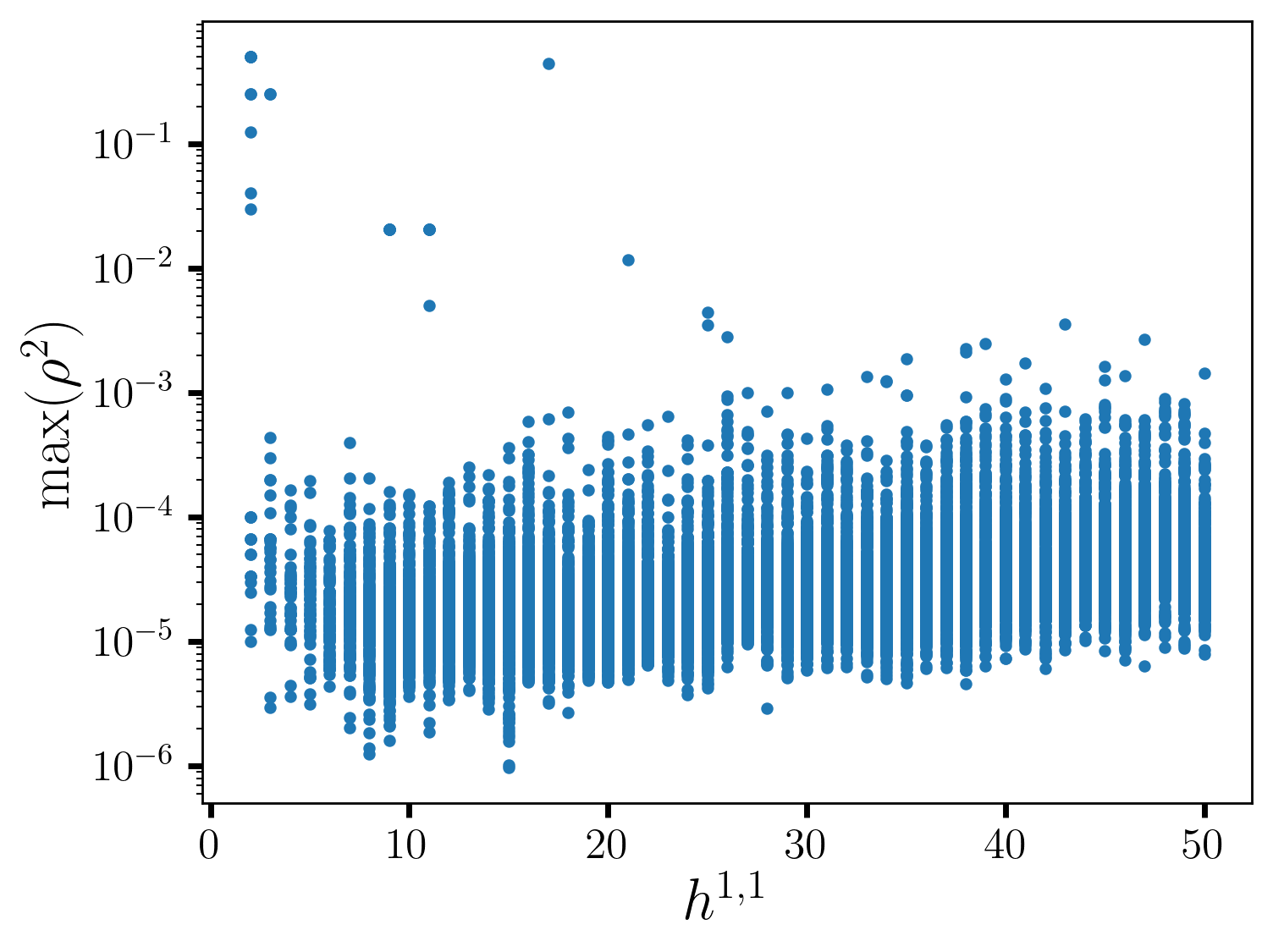}
			\caption{$C_2$; $f=f_{Q=q}$.}
		\end{subfigure}
		\begin{subfigure}{0.49\textwidth}
			\centering
			\includegraphics[width=\textwidth]{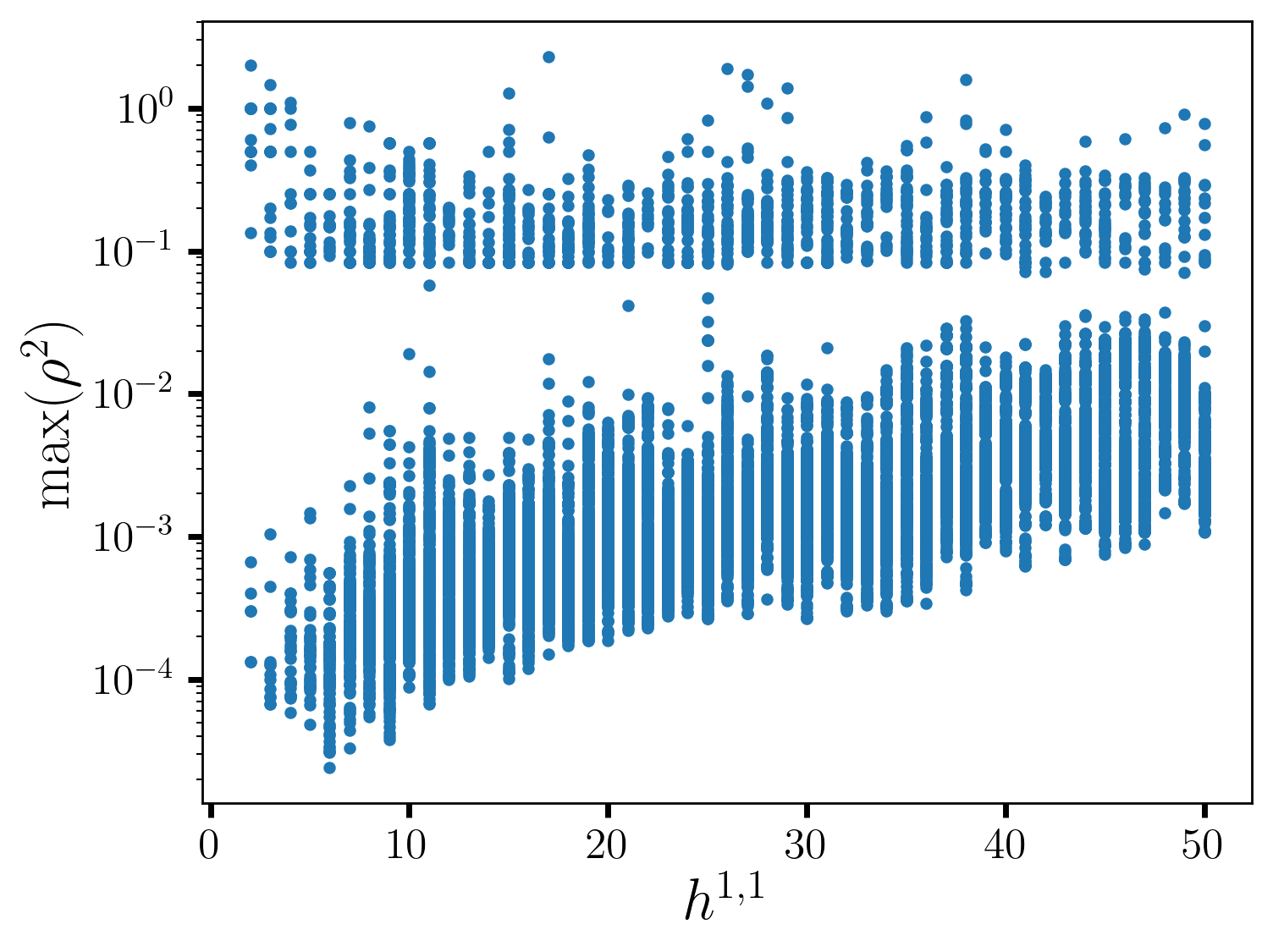}
			\caption{$C_4$; $f=f_{P=q}$.}
		\end{subfigure}
		\centering
		\begin{subfigure}{0.49\textwidth}
			\centering
			\includegraphics[width=\textwidth]{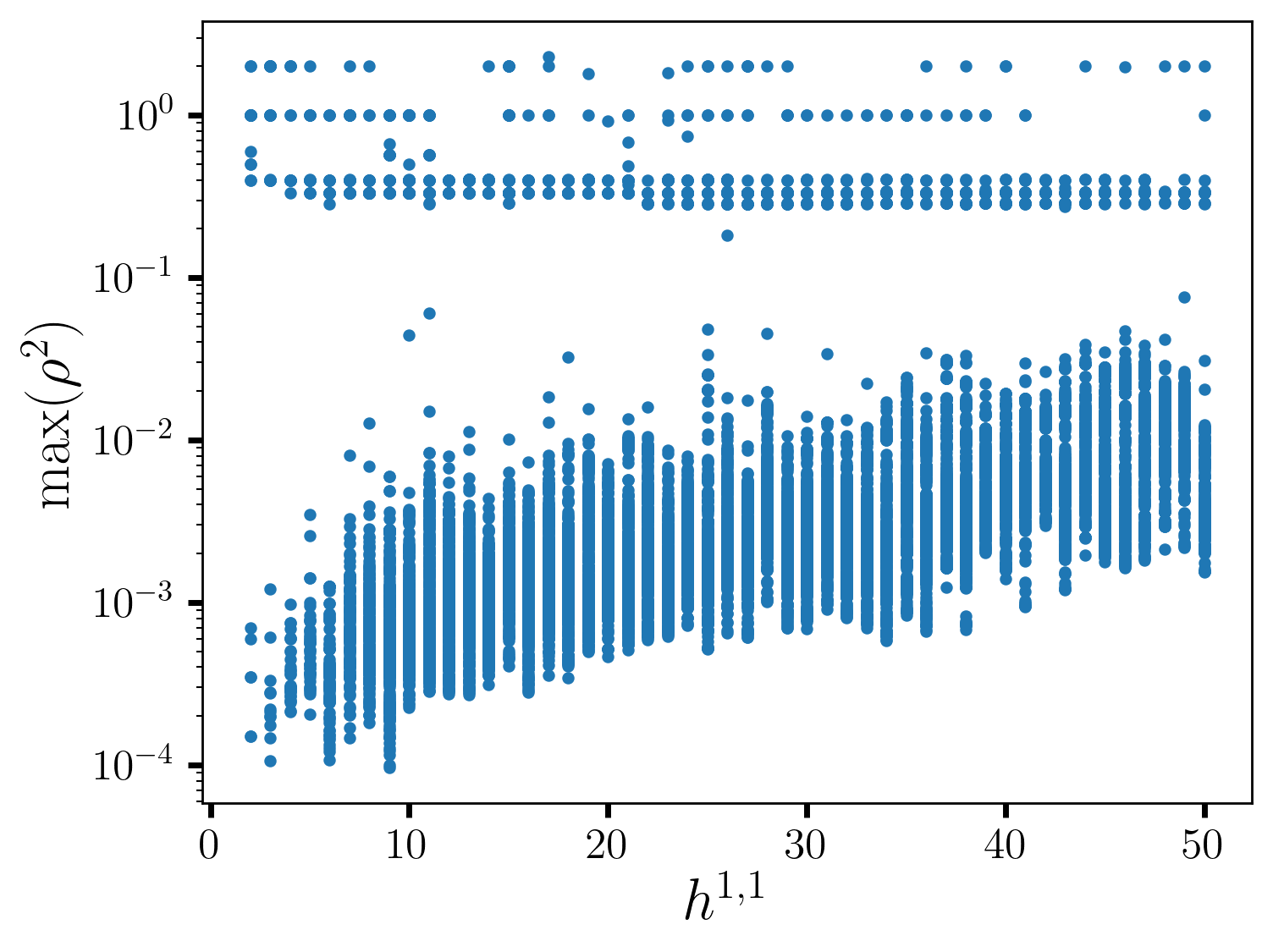}
			\caption{$C_4$; $f=f_{Q=\tildeq}$.}
		\end{subfigure}
		\caption{Plots of the maximum $\rho^2$, for different definitions of $\rho$, both $C_2$ and $C_4$ axions, where each point is near a facet. Note the logarithmic scale on the y-axis.}
		\label{plots.rho.facet.all.log.axis}
	\end{figure}
	
	\autoref{plots.rho.facet.all.log.axis} shows the plots of $\rho_{\rm max}$ near facets.  In total, 20,495 facets were scanned across 2116 Calabi-Yau manifolds.
	There were a further 47 Calabi-Yau geometries for which the algorithm failed to generate a point near any facets. We limited our scan to $h^{1,1}\leq50$ so as to avoid computational difficulties that arise when approaching facets for large $h^{1,1}$. 
    In the language of \secref{sec.analytics}, the algorithm took $L \approx 10^4$.

One striking feature of the $C_4$ axion plots is the separation of the data points into an upper band and lower band. These bands correspond, respectively, to the analytic scaling behaviors $\rho_{\rm max}^2 \sim L^0$ and $\rho_{\rm max}^2 \sim L^{-1}$, as seen in \autoref{table.scaling.C4}. In contrast, for the $C_2$ axions, the vast majority of the data points satisfy
$\rho_{\rm max}^2 \sim L^{-1}$. This indicates that the types of facets with $\alpha_{C_2} =0$ are very rare for large values of $h^{1,1}$.%
\enlargethispage{\baselineskip}%From {https://tex.stackexchange.com/questions/32208/footnote-runs-onto-second-page} and {https://texfaq.org/FAQ-splitfoot}; to get the footnote to not overflow to the next page, make this page one line longer.
\footnote{Indeed, type 3 codimension-1 boundaries of the K\"ahler cone with $\cV \sim s^2 L^3$ appear only for $h^{1,1}=2$. For large $h^{1,1}$, the majority of facets represent type 1 (flop) boundaries, and most of the remainder are type 2(a) boundaries.}

\section{Conclusions}
\label{sec:conclusions}

We have argued that extra-dimensional axions in Type IIB string theory compactifications obey the bound
\begin{equation}
    M_s \lesssim 2\pi \sqrt{S} f.
\end{equation}
Geometrically, this corresponds to a relationship between 2-cycle or 4-cycle volumes (which determine $S$), the overall Calabi-Yau volume (which determines $M_s$), and integrals of the form $\int \omega \wedge \star \omega$ for harmonic forms $\omega$ (which determine $f$). This underlying geometric relationship could be interpreted as a different physical bound by compactifying Type IIA string theory or M-theory on the same Calabi-Yau manifold (in a similar spirit to arguments in~\cite{Brown:2015iha, Hebecker:2015zss}). However, we are not aware of compelling physical implications of these reinterpreted bounds, so we will not present them in detail here.

When the bound~\eqref{stringscalebound} was originally proposed in~\cite{Reece:2025thc}, a significant emphasis was put on an argument based on the existence of axion strings with tension satisfying the co-scaling relation~\eqref{eq:axioncoscaling}. In view of our current knowledge, this was not the best approach: the bound survives, and indeed it is often satisfied by an even larger margin, in cases where co-scaling fails. The alternative arguments based on naturalness or unitarity may be the best motivation for the bound.

Several previous quantitative studies of the Type IIB axiverse have focused on the scaling of various quantities with $h^{1,1}$, which determines the number of axions~\cite{Demirtas:2018akl,Halverson:2019cmy,Mehta:2020kwu,Mehta:2021pwf,Demirtas:2021gsq,Gendler:2023kjt,Cheng:2025ggf,Fallon:2025lvn}. An important conclusion of such studies is that demanding that all cycles of the Calabi-Yau have volume at least of order-one in string units for calculational control requires the overall volume to become very large at large $h^{1,1}$, in turn leading to small string scale and small $f$. On the other hand, even at small $h^{1,1}$, it is possible that volumes are exponentially large and that $f$ is correspondingly small. This may happen due to an underlying moduli stabilization mechanism like LVS (the Large Volume Scenario)~\cite{Balasubramanian:2005zx}, and indeed, in that context $f$ is expected to be of order the string scale~\cite{Conlon:2006tq}. Our results are compatible with all of this past work but show that the relationship between $f$ and $M_s$ is quite general, not merely an artifact of focusing on large $h^{1,1}$, on the tip of the stretched K\"ahler cone, or on any details of a moduli-stabilization scenario. 

The Chern-Simons interaction strength of an axion (or its saxion partner) with gauge fields that have gauge coupling $g$ is of order $\frac{1}{f S} =\frac{g^2}{8\pi^2 f}$ (with $S$ the Yang-Mills instanton action). We can rearrange the bound~\eqref{stringscalebound} and combine it with the axion weak gravity  conjecture~\cite{Arkanihamed:2006dz,DiUbaldo:2026rly,Maldacena:2026jqd} to obtain a two-sided bound on this interaction strength:
\begin{equation}
\frac{1}{M_\mathrm{Pl}} \lesssim \frac{g^2}{8\pi^2 f} \lesssim \frac{g}{\sqrt{2} M_s}.    
\end{equation} 
Viewed in this way, the second $\lesssim$ relationship, which is a restatement of the quantum gravity cutoff from axions, seems very reasonable. $M_s$ is the cutoff scale at which local effective field theory breaks down, so it is natural that a fundamental interaction should be suppressed by $1/M_s$, and the coupling is further suppressed relative to this by at least one weak coupling factor $g$. In the special case that the string scale is suppressed relative to $M_\mathrm{Pl}$ by the same modulus that controls the coupling $g$, we expect that $M_s \sim g M_\mathrm{Pl}$ (at the WGC scale for the gauge theory~\cite{Arkanihamed:2006dz, Heidenreich:2021yda}). Then the two-sided bound degenerates (both $\lesssim$ become $\sim$) and the axion interacts with the same strength as gravity. This happens in the case of the heterotic string, for example. In such cases we expect the co-scaling relation~\eqref{eq:axioncoscaling} to be satisfied as well. More generally, it may happen that $M_s \ll g M_\mathrm{Pl}$, and the interactions of the axion and the saxion typically have a strength far above the Planck scale.

Our results strengthen the argument made in~\cite{Reece:2025thc} that the experimental discovery of an axion can give us an important clue about the fundamental cutoff scale of quantum gravity. The important caveat is that the argument assumes the axion is an extra-dimensional axion. A conventional 4d axion would lead to precisely the opposite conclusion: $f$ is a scale at which 4d physics applies, possibly many orders of magnitude below the fundamental cutoff. It is important, then, to continue to search for experimental or observational signatures that might clarify if an axion is an extra-dimensional axion or a 4d axion. The former seems strongly motivated both from the bottom-up axion quality problem and from top-down string theory constructions, but an unambiguous empirical discrimination between the two cases is elusive.

\section*{Acknowledgments}

We thank Josh Benabou, Katie Fraser, Naomi Gendler, and Ben Safdi for useful discussions.
MR is supported in part by the DOE Grant DE-SC0013607. This work was performed in part at the Aspen Center for Physics, which is supported by National Science Foundation grant PHY-2210452.
The work of TR was supported in part by STFC through grant ST/T000708/1 and by the Royal Society through grant RGS/R2/252603.
The work of CT was supported by a studentship from STFC through grant ST/Y509334/1 and by the Royal Society through grant RGS/R2/252603.
    
\appendix

\section{Universal patterns at the tip of the stretched K\"ahler cone at large \texorpdfstring{$h^{1,1}$}{h1,1}} \label{sec.structure.of.tip.plots}

As shown above in \autoref{plots.rho.tip.all.linear.axis}, our numerical sweep of the Calabi-Yau landscape found many points at large $h^{1,1}$ with
\begin{equation}
\brap{\rhoAx{2}{\max}}^2 \approx \frac13\,,~~~~\brap{\rhoInst{4}{\max}}^2 \approx 6\,,
\label{obseq1}
\end{equation}
as well as an apparent lower bound
\begin{equation}
\brap{\rhoInst{2}{\max}}^2 \gtrsim 3\,.
\label{obseq2}
\end{equation}
In this appendix, we provide a (partial) explanations for these observed patterns.
    
	 To begin, we recall that the tip of the stretched K\"ahler cone is the point closest to the origin that is at least distance $1$ away from all facets.
     However, when the K\"ahler cone is non-simplicial (which occurs with probability $p \approx 1$ for large $h^{1,1}$), it is generically impossible to find a point that has distance 1 from every facet.
    Instead, for large $h^{1,1}$, the tip of the stretched K\"ahler cone will have a wide spread of distances from the different facets -- starting at $1$ and extending all the way to some large $L$ -- and so we are in a regime where we can attempt to apply the analytic results of \S\ref{sec.analytics}.

    Next, let us define the $\cS$-map as \cite{Reece:2025zva}:
	\begin{align}
		\cS: \cK_\text{hyp} \times H_4\brap{X,\bbR} \rightarrow &H_2\brap{X,\bbR}\\
		(t^i,{\tilde q}^j) \mapsto &\cF_{ij}\brap{t} {\tilde q}^j\,.
	\end{align}
    We then consider a pair of charges $\tilde q\in\cEker$ and $q=\mathcal{S}(\tilde q) =  F\brap{t} \cdot \tilde q$, where $F$ is the matrix given by $\cF_{ij}$.

    This pair of charges then satisfies:
    \begin{align}
        q \cdot t
            &= 2 \tilde q \cdot \tau \sim L^0\,,~~~~~\text{ ($\tilde q \in \cEker$)}\\
        \tilde q \cdot q
            &= \tilde q \cdot F \cdot \tilde q \sim L^0\,,~~~~~\text{ ($\tilde q \in \cEker$)}\\
        \tilde q \cdot \metric \cdot \tilde q
            &= - \tilde q \cdot F \cdot \tilde q  + \frac{\brap{2\tilde q \cdot \tau}^2}{\Vol} \\
            &= - \tilde q \cdot q + \cO\brap{\frac1\Vol}\,,\\
        q \cdot \metricInvNoIndex\cdot q
            &= -  q \cdot F^{-1} \cdot  q + \frac{\brap{ q \cdot t}^2}{2\Vol}\\
            &= -  \tilde q \cdot F \cdot \tilde q + \frac{\brap{ q \cdot t}^2}{2\Vol}\\
            &= - \tilde q \cdot q + \cO\brap{\frac1\Vol}\,,
    \end{align}
    where $\frac1\Vol \ll L^{0}$.

    If $q \in H_2(X, \mathbb{Z})$ is effective, then $q \cdot t \geq1$ for $t$ in the stretched K\"ahler cone. If this inequality is approximately saturated, then $\tilde q \cdot \tau \approx \frac12$, by definition of the $\mathcal{S}$-map. If $\tilde q$ is also effective, then
    \begin{align}
        \brap{\rhoInst{4}{\max}}^2 &\gtrsim 2\brap{- \tilde q \cdot q} \label{c4seq}\\
        \brap{\rhoAx{2}{\max}}^2 &\gtrsim \frac1{\brap{-\tilde q \cdot q}}\\
        \brap{\rhoInst{2}{\max}}^2 &\gtrsim \brap{- \tilde q \cdot q}\,. \label{c2seq}
    \end{align}

    Upon investigating a handful of examples with $\brap{\rhoInst{4}{\max}}^2\approx 6$ (some with and some without $\brap{\rhoInst{2}{\max}}^2 \approx 3$), in each such example we find at least one $\tilde q^j = \delta^j_{i^\ast}\in\cEker$ that has $\tau_{i^\ast}\approx\frac12$, with $\cS(\tilde q) \approx q$, for $q$ a generator of the Mori cone satisfying $q \cdot t\approx1$. In the examples considered, this pair of charges $q, \tilde q$ have $- \tilde q \cdot q=3$. By \eqref{c4seq}-\eqref{c2seq}, this implies
    \begin{align}
        \brap{\rhoInst{4}{\max}}^2 &\gtrsim 6 \label{c4instbound}\\
        \brap{\rhoAx{2}{\max}}^2 &\gtrsim \frac13 \label{c2axbound}\\
        \brap{\rhoInst{2}{\max}}^2 &\gtrsim 3 \,, \label{c2instbound}
    \end{align}
    which matches the values observed in \eqref{obseq1}-\eqref{obseq2}.

    This explanation leaves a handful of open questions, however. The Dirac pairing $- \tilde q \cdot q$ is required by charge quantization to be an integer, but it is unclear why this large class of examples has $- \tilde q \cdot q = 3$. It is also unclear to us why $\rhoInst{4}{\max}$ and $\rhoAx{2}{\max}$ nearly always saturate the bounds in \eqref{c4instbound}-\eqref{c2axbound}, whereas \eqref{c2instbound} is often satisfied with room to spare. We leave these questions for future work.

\bibliographystyle{utphys}
\bibliography{ref}

@article{Raman:2024fcv,
    author = "Raman, Sanjay and Vafa, Cumrun",
    title = "{Swampland and the Geometry of Marked Moduli Spaces}",
    eprint = "2405.11611",
    archivePrefix = "arXiv",
    primaryClass = "hep-th",
    month = "5",
    year = "2024"
}

@article{Witten:1996qb,
	Archiveprefix = {arXiv},
	Author = {Witten, Edward},
	Doi = {10.1016/0550-3213(96)00212-X},
	Eprint = {hep-th/9603150},
	Journal = {Nucl. Phys.},
	Pages = {195-216},
	Primaryclass = {hep-th},
	Reportnumber = {IASSNS-HEP-96-26},
	Slaccitation = {%%CITATION = HEP-TH/9603150;%%},
	Title = {{Phase transitions in M theory and F theory}},
	Volume = {B471},
	Year = {1996}}

@article{Gendler:2022ztv,
    author = "Gendler, Naomi and Heidenreich, Ben and McAllister, Liam and Moritz, Jakob and Rudelius, Tom",
    title = "{Moduli space reconstruction and Weak Gravity}",
    eprint = "2212.10573",
    archivePrefix = "arXiv",
    primaryClass = "hep-th",
    reportNumber = "ACFI-T22-10",
    doi = "10.1007/JHEP12(2023)134",
    journal = "JHEP",
    volume = "12",
    pages = "134",
    year = "2023"
}

@article{Kreuzer:2000xy,
	Archiveprefix = {arXiv},
	Author = {Kreuzer, Maximilian and Skarke, Harald},
	Doi = {10.4310/ATMP.2000.v4.n6.a2},
	Eprint = {hep-th/0002240},
	Journal = {Adv. Theor. Math. Phys.},
	Pages = {1209--1230},
	Reportnumber = {HUB-EP-00-13, TUW-00-07},
	Title = {{Complete classification of reflexive polyhedra in four-dimensions}},
	Volume = {4},
	Year = {2000}}

@article{Harlow:2022ich,
    author = "Harlow, Daniel and Heidenreich, Ben and Reece, Matthew and Rudelius, Tom",
    title = "{Weak gravity conjecture}",
    eprint = "2201.08380",
    archivePrefix = "arXiv",
    primaryClass = "hep-th",
    reportNumber = "ACFI-T22-01",
    doi = "10.1103/RevModPhys.95.035003",
    journal = "Rev. Mod. Phys.",
    volume = "95",
    number = "3",
    pages = "035003",
    year = "2023"
}

@article{rudelius:2015xta,
	Archiveprefix = {arXiv},
	Author = {Rudelius, Tom},
	Doi = {10.1088/1475-7516/2015/9/020},
	Eprint = {1503.00795},
	Journal = {JCAP},
	Pages = {020},
	Primaryclass = {hep-th},
	Slaccitation = {%%CITATION = ARXIV:1503.00795;%%},
	Title = {{Constraints on Axion Inflation from the Weak Gravity Conjecture}},
	Volume = {09},
	Year = {2015}}

@article{Heidenreich:2021yda,
    author = "Heidenreich, Ben and Reece, Matthew and Rudelius, Tom",
    title = "{The Weak Gravity Conjecture and axion strings}",
    eprint = "2108.11383",
    archivePrefix = "arXiv",
    primaryClass = "hep-th",
    reportNumber = "ACFI-T21-10",
    doi = "10.1007/JHEP11(2021)004",
    journal = "JHEP",
    volume = "11",
    pages = "004",
    year = "2021"
}

@article{Lanza:2020qmt,
	Archiveprefix = {arXiv},
	Author = {Lanza, Stefano and Marchesano, Fernando and Martucci, Luca and Valenzuela, Irene},
	Doi = {10.1007/JHEP02(2021)006},
	Eprint = {2006.15154},
	Journal = {JHEP},
	Pages = {006},
	Primaryclass = {hep-th},
	Title = {{Swampland Conjectures for Strings and Membranes}},
	Volume = {02},
	Year = {2021}}

@article{Alim:2021vhs,
    author = "Alim, Murad and Heidenreich, Ben and Rudelius, Tom",
    title = "{The Weak Gravity Conjecture and BPS Particles}",
    eprint = "2108.08309",
    archivePrefix = "arXiv",
    primaryClass = "hep-th",
    reportNumber = "ACFI-T21-09",
    doi = "10.1002/prop.202100125",
    journal = "Fortsch. Phys.",
    volume = "69",
    number = "11-12",
    pages = "2100125",
    year = "2021"
}

@article{Brown:2015iha,
	Archiveprefix = {arXiv},
	Author = {Brown, Jon and Cottrell, William and Shiu, Gary and Soler, Pablo},
	Doi = {10.1007/JHEP10(2015)023},
	Eprint = {1503.04783},
	Journal = {JHEP},
	Pages = {023},
	Primaryclass = {hep-th},
	Reportnumber = {MAD-TH-15-04},
	Title = {{Fencing in the Swampland: Quantum Gravity Constraints on Large Field Inflation}},
	Volume = {10},
	Year = {2015}}

@article{Lee:2019wij,
    author = "Lee, Seung-Joo and Lerche, Wolfgang and Weigand, Timo",
    title = "{Emergent strings from infinite distance limits}",
    eprint = "1910.01135",
    archivePrefix = "arXiv",
    primaryClass = "hep-th",
    reportNumber = "CERN-TH-2019-159",
    doi = "10.1007/JHEP02(2022)190",
    journal = "JHEP",
    volume = "02",
    pages = "190",
    year = "2022"
}

@article{Reece:2018zvv,
	Archiveprefix = {arXiv},
	Author = {Reece, Matthew},
	Doi = {10.1007/JHEP07(2019)181},
	Eprint = {1808.09966},
	Journal = {JHEP},
	Pages = {181},
	Primaryclass = {hep-th},
	Title = {{Photon Masses in the Landscape and the Swampland}},
	Volume = {07},
	Year = {2019}}

@article{Ooguri:2006in,
	Archiveprefix = {arXiv},
	Author = {Ooguri, Hirosi and Vafa, Cumrun},
	Doi = {10.1016/j.nuclphysb.2006.10.033},
	Eprint = {hep-th/0605264},
	Journal = {Nucl.Phys.},
	Pages = {21-33},
	Primaryclass = {hep-th},
	Reportnumber = {CALT-68-2600, HUTP-06-A017},
	Slaccitation = {%%CITATION = HEP-TH/0605264;%%},
	Title = {{On the Geometry of the String Landscape and the Swampland}},
	Volume = {B766},
	Year = {2007}}

@article{Arkanihamed:2006dz,
	Archiveprefix = {arXiv},
	Author = {Arkani-Hamed, Nima and Motl, Lubos and Nicolis, Alberto and Vafa, Cumrun},
	Doi = {10.1088/1126-6708/2007/06/060},
	Eprint = {hep-th/0601001},
	Journal = {JHEP},
	Pages = {060},
	Primaryclass = {hep-th},
	Reportnumber = {HUTP-05-A0057},
	Slaccitation = {%%CITATION = HEP-TH/0601001;%%},
	Title = {{The String landscape, black holes and gravity as the weakest force}},
	Volume = {0706},
	Year = {2007}}

@article{Conlon:2006tq,
	Archiveprefix = {arXiv},
	Author = {Conlon, Joseph P.},
	Doi = {10.1088/1126-6708/2006/05/078},
	Eprint = {hep-th/0602233},
	Journal = {JHEP},
	Pages = {078},
	Primaryclass = {hep-th},
	Reportnumber = {DAMTP-2006-17},
	Slaccitation = {%%CITATION = HEP-TH/0602233;%%},
	Title = {{The QCD axion and moduli stabilisation}},
	Volume = {05},
	Year = {2006}}

@article{Svrcek:2006yi,
	Archiveprefix = {arXiv},
	Author = {Svrcek, Peter and Witten, Edward},
	Doi = {10.1088/1126-6708/2006/06/051},
	Eprint = {hep-th/0605206},
	Journal = {JHEP},
	Pages = {051},
	Primaryclass = {hep-th},
	Reportnumber = {SLAC-PUB-11894},
	Slaccitation = {%%CITATION = HEP-TH/0605206;%%},
	Title = {{Axions In String Theory}},
	Volume = {06},
	Year = {2006}}

@article{Witten:1984dg,
	Author = {Witten, Edward},
	Doi = {10.1016/0370-2693(84)90422-2},
	Journal = {Phys. Lett.},
	Pages = {351-356},
	Reportnumber = {Print-84-0838 (PRINCETON)},
	Slaccitation = {%%CITATION = PHLTA,149B,351;%%},
	Title = {{Some Properties of O(32) Superstrings}},
	Volume = {149B},
	Year = {1984}}

@article{Barr:1985hk,
	Author = {Barr, Stephen M.},
	Doi = {10.1016/0370-2693(85)90440-X},
	Journal = {Phys. Lett.},
	Pages = {397-400},
	Reportnumber = {UW-ER-40048-07-P5},
	Slaccitation = {%%CITATION = PHLTA,158B,397;%%},
	Title = {{Harmless Axions in Superstring Theories}},
	Volume = {158B},
	Year = {1985}}

@article{Choi:1985je,
	Author = {Choi, Kiwoon and Kim, Jihn E.},
	Doi = {10.1016/0370-2693(85)90416-2},
	Journal = {Phys. Lett.},
	Note = {[Erratum: Phys. Lett.156B,452(1985)]},
	Pages = {393},
	Reportnumber = {HUTP-85/A013},
	Slaccitation = {%%CITATION = PHLTA,154B,393;%%},
	Title = {{Harmful Axions in Superstring Models}},
	Volume = {154B},
	Year = {1985}}

@misc{BPSstrings,
	Author = {Heidenreich, Ben and Pittman, Nicholas and Rudelius, Tom},
	Note = {to appear},
	Title = {The {Weak Gravity Conjecture and BPS Strings}},
	Year = {2026}}

@article{Brodie:2021ain,
    author = "Brodie, Callum R. and Constantin, Andrei and Lukas, Andre and Ruehle, Fabian",
    title = "{Swampland conjectures and infinite flop chains}",
    eprint = "2104.03325",
    archivePrefix = "arXiv",
    primaryClass = "hep-th",
    reportNumber = "CERN-TH-2021-051",
    doi = "10.1103/PhysRevD.104.046008",
    journal = "Phys. Rev. D",
    volume = "104",
    number = "4",
    pages = "046008",
    year = "2021"
}

@article{Marchesano:2023thx,
    author = "Marchesano, Fernando and Melotti, Luca and Paoloni, Lorenzo",
    title = "{On the moduli space curvature at infinity}",
    eprint = "2311.07979",
    archivePrefix = "arXiv",
    primaryClass = "hep-th",
    doi = "10.1007/JHEP02(2024)103",
    journal = "JHEP",
    volume = "02",
    pages = "103",
    year = "2024"
}

@article{Marchesano:2024tod,
    author = "Marchesano, Fernando and Melotti, Luca and Wiesner, Max",
    title = "{Asymptotic curvature divergences and non-gravitational theories}",
    eprint = "2409.02991",
    archivePrefix = "arXiv",
    primaryClass = "hep-th",
    doi = "10.1007/JHEP02(2025)151",
    journal = "JHEP",
    volume = "02",
    pages = "151",
    year = "2025"
}

@article{Castellano:2024gwi,
    author = "Castellano, Alberto and Marchesano, Fernando and Melotti, Luca and Paoloni, Lorenzo",
    title = "{The Moduli Space Curvature and the Weak Gravity Conjecture}",
    eprint = "2410.10966",
    archivePrefix = "arXiv",
    primaryClass = "hep-th",
    month = "10",
    year = "2024"
}

@article{Blanco:2025qom,
    author = "Blanco, Alejandro and Marchesano, Fernando and Melotti, Luca",
    title = "{Curvature divergences in 5d $\mathcal{N}=1$ supergravity}",
    eprint = "2505.05558",
    archivePrefix = "arXiv",
    primaryClass = "hep-th",
    doi = "10.1007/JHEP11(2025)026",
    journal = "JHEP",
    volume = "11",
    pages = "026",
    year = "2025"
}

@article{Veneziano:2001ah,
    author = "Veneziano, G.",
    title = "{Large N bounds on, and compositeness limit of, gauge and gravitational interactions}",
    eprint = "hep-th/0110129",
    archivePrefix = "arXiv",
    reportNumber = "CERN-TH-2001-278",
    doi = "10.1088/1126-6708/2002/06/051",
    journal = "JHEP",
    volume = "06",
    pages = "051",
    year = "2002"
}

@article{Balasubramanian:2005zx,
    author = "Balasubramanian, Vijay and Berglund, Per and Conlon, Joseph P. and Quevedo, Fernando",
    title = "{Systematics of moduli stabilisation in Calabi-Yau flux compactifications}",
    eprint = "hep-th/0502058",
    archivePrefix = "arXiv",
    reportNumber = "DAMTP-2005-10, UNH-05-01, UPR-1109-T",
    doi = "10.1088/1126-6708/2005/03/007",
    journal = "JHEP",
    volume = "03",
    pages = "007",
    year = "2005"
}

@article{Choi:2003wr,
    author = "Choi, Ki-woon",
    title = "{A QCD axion from higher dimensional gauge field}",
    eprint = "hep-ph/0308024",
    archivePrefix = "arXiv",
    reportNumber = "KAIST-TH-2003-07",
    doi = "10.1103/PhysRevLett.92.101602",
    journal = "Phys. Rev. Lett.",
    volume = "92",
    pages = "101602",
    year = "2004"
}

@article{Lanza:2022zyg,
    author = "Lanza, Stefano and Marchesano, Fernando and Martucci, Luca and Valenzuela, Irene",
    title = "{Large Field Distances from EFT strings}",
    eprint = "2205.04532",
    archivePrefix = "arXiv",
    primaryClass = "hep-th",
    doi = "10.22323/1.406.0169",
    journal = "PoS",
    volume = "CORFU2021",
    pages = "169",
    year = "2022"
}

@article{Hebecker:2015zss,
    author = "Hebecker, Arthur and Rompineve, Fabrizio and Westphal, Alexander",
    title = "{Axion Monodromy and the Weak Gravity Conjecture}",
    eprint = "1512.03768",
    archivePrefix = "arXiv",
    primaryClass = "hep-th",
    reportNumber = "DESY-15-242",
    doi = "10.1007/JHEP04(2016)157",
    journal = "JHEP",
    volume = "04",
    pages = "157",
    year = "2016"
}

@article{Seo:2024zzs,
    author = "Seo, Min-Seok",
    title = "{Axion species scale and axion weak gravity conjecture-like bound}",
    eprint = "2407.16156",
    archivePrefix = "arXiv",
    primaryClass = "hep-th",
    doi = "10.1007/JHEP11(2024)082",
    journal = "JHEP",
    volume = "11",
    pages = "082",
    year = "2024"
}

@article{Grimm:2007hs,
    author = "Grimm, Thomas W.",
    title = "{Axion inflation in type II string theory}",
    eprint = "0710.3883",
    archivePrefix = "arXiv",
    primaryClass = "hep-th",
    reportNumber = "BONN-TH-2007-11, MAD-TH-07-11",
    doi = "10.1103/PhysRevD.77.126007",
    journal = "Phys. Rev. D",
    volume = "77",
    pages = "126007",
    year = "2008"
}

@article{Benabou:2025kgx,
    author = "Benabou, Joshua N. and Fraser, Katherine and Reig, Mario and Safdi, Benjamin R.",
    title = "{String theory and grand unification suggest a submicroelectronvolt QCD axion}",
    eprint = "2505.15884",
    archivePrefix = "arXiv",
    primaryClass = "hep-ph",
    doi = "10.1103/lthr-97lm",
    journal = "Phys. Rev. D",
    volume = "112",
    number = "6",
    pages = "066003",
    year = "2025"
}

@article{Reece:2025zva,
    author = "Reece, Matthew and Rudelius, Tom and Tudball, Christopher",
    title = "{Co-scaling and alignment of electric and magnetic towers}",
    eprint = "2505.22713",
    archivePrefix = "arXiv",
    primaryClass = "hep-th",
    doi = "10.1007/JHEP09(2025)146",
    journal = "JHEP",
    volume = "09",
    pages = "146",
    year = "2025"
}

@article{Craig:2024dnl,
    author = "Craig, Nathaniel and Kongsore, Marius",
    title = "{High-quality axions from higher-form symmetries in extra dimensions}",
    eprint = "2408.10295",
    archivePrefix = "arXiv",
    primaryClass = "hep-ph",
    doi = "10.1103/PhysRevD.111.015047",
    journal = "Phys. Rev. D",
    volume = "111",
    number = "1",
    pages = "015047",
    year = "2025"
}

@article{Demirtas:2018akl,
    author = "Demirtas, Mehmet and Long, Cody and McAllister, Liam and Stillman, Mike",
    title = "{The Kreuzer-Skarke Axiverse}",
    eprint = "1808.01282",
    archivePrefix = "arXiv",
    primaryClass = "hep-th",
    doi = "10.1007/JHEP04(2020)138",
    journal = "JHEP",
    volume = "04",
    pages = "138",
    year = "2020"
}

@article{Cicoli:2012sz,
    author = "Cicoli, Michele and Goodsell, Mark and Ringwald, Andreas",
    title = "{The type IIB string axiverse and its low-energy phenomenology}",
    eprint = "1206.0819",
    archivePrefix = "arXiv",
    primaryClass = "hep-th",
    reportNumber = "DESY-12-058, CERN-PH-TH-2012-153",
    doi = "10.1007/JHEP10(2012)146",
    journal = "JHEP",
    volume = "10",
    pages = "146",
    year = "2012"
}

@article{Acharya:2010zx,
    author = "Acharya, Bobby Samir and Bobkov, Konstantin and Kumar, Piyush",
    title = "{An M Theory Solution to the Strong CP Problem and Constraints on the Axiverse}",
    eprint = "1004.5138",
    archivePrefix = "arXiv",
    primaryClass = "hep-th",
    doi = "10.1007/JHEP11(2010)105",
    journal = "JHEP",
    volume = "11",
    pages = "105",
    year = "2010"
}

@article{Demirtas:2021gsq,
    author = "Demirtas, Mehmet and Gendler, Naomi and Long, Cody and McAllister, Liam and Moritz, Jakob",
    title = "{PQ axiverse}",
    eprint = "2112.04503",
    archivePrefix = "arXiv",
    primaryClass = "hep-th",
    doi = "10.1007/JHEP06(2023)092",
    journal = "JHEP",
    volume = "06",
    pages = "092",
    year = "2023"
}

@article{Gendler:2023kjt,
    author = "Gendler, Naomi and Marsh, David J. E. and McAllister, Liam and Moritz, Jakob",
    title = "{Glimmers from the axiverse}",
    eprint = "2309.13145",
    archivePrefix = "arXiv",
    primaryClass = "hep-th",
    reportNumber = "KCL-PH-TH/2023-49",
    doi = "10.1088/1475-7516/2024/09/071",
    journal = "JCAP",
    volume = "09",
    pages = "071",
    year = "2024"
}

@article{Fallon:2025lvn,
    author = "Fallon, Sebastian Vander Ploeg and Halverson, James and McAllister, Liam and Zhu, Yunhao",
    title = "{F-theory Axiverse}",
    eprint = "2511.20458",
    archivePrefix = "arXiv",
    primaryClass = "hep-th",
    month = "11",
    year = "2025"
}

@article{Brivio:2021fog,
    author = "Brivio, I. and {\'E}boli, O. J. P. and Gonzalez-Garcia, M. C.",
    title = "{Unitarity constraints on ALP interactions}",
    eprint = "2106.05977",
    archivePrefix = "arXiv",
    primaryClass = "hep-ph",
    reportNumber = "YITP-SB-2021-8",
    doi = "10.1103/PhysRevD.104.035027",
    journal = "Phys. Rev. D",
    volume = "104",
    number = "3",
    pages = "035027",
    year = "2021"
}

@article{Dine:1986bg,
    author = "Dine, Michael and Seiberg, Nathan",
    title = "{String Theory and the Strong {CP} Problem}",
    reportNumber = "Print-86-0091 (CITY COLL.,N.Y.), CCNY-HEP-86/2",
    doi = "10.1016/0550-3213(86)90043-X",
    journal = "Nucl. Phys. B",
    volume = "273",
    pages = "109--124",
    year = "1986"
}

@article{Banks:1996ea,
    author = "Banks, Tom and Dine, Michael",
    title = "{The Cosmology of string theoretic axions}",
    eprint = "hep-th/9608197",
    archivePrefix = "arXiv",
    reportNumber = "SCIPP-96-31, RU-96-95A, RU-96-95",
    doi = "10.1016/S0550-3213(97)00413-6",
    journal = "Nucl. Phys. B",
    volume = "505",
    pages = "445--460",
    year = "1997"
}

@article{Arvanitaki:2009fg,
    author = "Arvanitaki, Asimina and Dimopoulos, Savas and Dubovsky, Sergei and Kaloper, Nemanja and March-Russell, John",
    title = "{String Axiverse}",
    eprint = "0905.4720",
    archivePrefix = "arXiv",
    primaryClass = "hep-th",
    doi = "10.1103/PhysRevD.81.123530",
    journal = "Phys. Rev. D",
    volume = "81",
    pages = "123530",
    year = "2010"
}

@article{Reece:2023czb,
    author = "Reece, Matthew",
    title = "{TASI Lectures: (No) Global Symmetries to Axion Physics}",
    eprint = "2304.08512",
    archivePrefix = "arXiv",
    primaryClass = "hep-ph",
    doi = "10.22323/1.439.0008",
    journal = "PoS",
    volume = "TASI2022",
    pages = "008",
    year = "2024"
}

@article{Reece:2025thc,
    author = "Reece, Matthew",
    title = "{Extra-dimensional axion expectations}",
    eprint = "2406.08543",
    archivePrefix = "arXiv",
    primaryClass = "hep-ph",
    doi = "10.1007/JHEP07(2025)130",
    journal = "JHEP",
    volume = "07",
    pages = "130",
    year = "2025"
}

@article{Brandt:2026wir,
    author = "Brandt, Bastian B. and Endr{\H{o}}di, Gergely and Hern{\'a}ndez Hern{\'a}ndez, Jos{\'e} Javier and Mark{\'o}, Gergely and Pannullo, Laurin",
    title = "{The axion-photon coupling from lattice Quantum Chromodynamics}",
    eprint = "2603.29153",
    archivePrefix = "arXiv",
    primaryClass = "hep-lat",
    month = "3",
    year = "2026"
}

@article{Fraser:2019ojt,
    author = "Fraser, Katherine and Reece, Matthew",
    title = "{Axion Periodicity and Coupling Quantization in the Presence of Mixing}",
    eprint = "1910.11349",
    archivePrefix = "arXiv",
    primaryClass = "hep-ph",
    doi = "10.1007/JHEP05(2020)066",
    journal = "JHEP",
    volume = "05",
    pages = "066",
    year = "2020"
}

@article{Mehta:2021pwf,
    author = "Mehta, Viraf M. and Demirtas, Mehmet and Long, Cody and Marsh, David J. E. and McAllister, Liam and Stott, Matthew J.",
    title = "{Superradiance in string theory}",
    eprint = "2103.06812",
    archivePrefix = "arXiv",
    primaryClass = "hep-th",
    doi = "10.1088/1475-7516/2021/07/033",
    journal = "JCAP",
    volume = "07",
    pages = "033",
    year = "2021"
}

@article{Kaplan:1985dv,
    author = "Kaplan, David B.",
    title = "{Opening the Axion Window}",
    reportNumber = "HUTP-85/A014",
    doi = "10.1016/0550-3213(85)90319-0",
    journal = "Nucl. Phys. B",
    volume = "260",
    pages = "215--226",
    year = "1985"
}

@article{Srednicki:1985xd,
    author = "Srednicki, Mark",
    title = "{Axion Couplings to Matter. 1. CP Conserving Parts}",
    reportNumber = "Print-85-0247 (UC,SANTA BARBARA)",
    doi = "10.1016/0550-3213(85)90054-9",
    journal = "Nucl. Phys. B",
    volume = "260",
    pages = "689--700",
    year = "1985"
}

@article{Georgi:1986df,
    author = "Georgi, Howard and Kaplan, David B. and Randall, Lisa",
    title = "{Manifesting the Invisible Axion at Low-energies}",
    reportNumber = "HUTP-86/A004",
    doi = "10.1016/0370-2693(86)90688-X",
    journal = "Phys. Lett. B",
    volume = "169",
    pages = "73--78",
    year = "1986"
}

@article{GrillidiCortona:2015jxo,
    author = "Grilli di Cortona, Giovanni and Hardy, Edward and Pardo Vega, Javier and Villadoro, Giovanni",
    title = "{The QCD axion, precisely}",
    eprint = "1511.02867",
    archivePrefix = "arXiv",
    primaryClass = "hep-ph",
    doi = "10.1007/JHEP01(2016)034",
    journal = "JHEP",
    volume = "01",
    pages = "034",
    year = "2016"
}

@article{Lu:2020rhp,
    author = "Lu, Zhen-Yan and Du, Meng-Lin and Guo, Feng-Kun and Mei{\ss}ner, Ulf-G. and Vonk, Thomas",
    title = "{QCD $\theta$-vacuum energy and axion properties}",
    eprint = "2003.01625",
    archivePrefix = "arXiv",
    primaryClass = "hep-ph",
    doi = "10.1007/JHEP05(2020)001",
    journal = "JHEP",
    volume = "05",
    pages = "001",
    year = "2020"
}

@article{Gao:2022xqz,
    author = "Gao, Rui and Guo, Zhi-Hui and Oller, J. A. and Zhou, Hai-Qing",
    title = "{Axion-meson mixing in light of recent lattice {\ensuremath{\eta}}{\textendash}{\ensuremath{\eta}}' simulations and their two-photon couplings within U(3) chiral theory}",
    eprint = "2211.02867",
    archivePrefix = "arXiv",
    primaryClass = "hep-ph",
    doi = "10.1007/JHEP04(2023)022",
    journal = "JHEP",
    volume = "04",
    pages = "022",
    year = "2023"
}

@article{Meggiolaro:2025yiu,
    author = "Meggiolaro, Enrico and Tamburini, Mirko",
    title = "{New study of the interactions of the axion with mesons and photons using a chiral effective Lagrangian model}",
    eprint = "2502.13615",
    archivePrefix = "arXiv",
    primaryClass = "hep-ph",
    reportNumber = "IFUP-TH/2025",
    doi = "10.1103/PhysRevD.111.095024",
    journal = "Phys. Rev. D",
    volume = "111",
    number = "9",
    pages = "095024",
    year = "2025"
}

@article{Halverson:2019kna,
    author = "Halverson, James and Long, Cody and Nelson, Brent and Salinas, Gustavo",
    title = "{Axion reheating in the string landscape}",
    eprint = "1903.04495",
    archivePrefix = "arXiv",
    primaryClass = "hep-th",
    doi = "10.1103/PhysRevD.99.086014",
    journal = "Phys. Rev. D",
    volume = "99",
    number = "8",
    pages = "086014",
    year = "2019"
}

@article{DiVecchia:1980yfw,
    author = "Di Vecchia, P. and Veneziano, G.",
    title = "{Chiral Dynamics in the Large n Limit}",
    reportNumber = "CERN-TH-2814",
    doi = "10.1016/0550-3213(80)90370-3",
    journal = "Nucl. Phys. B",
    volume = "171",
    pages = "253--272",
    year = "1980"
}

@article{smith1862systems,
  title={On systems of linear indeterminate equations and congruences},
  author={Smith, Henry John Stephen},
  journal={Proceedings of the Royal Society of London},
  number={11},
  pages={86--89},
  year={1862},
  publisher={The Royal Society London},
  doi="10.1098/rstl.1861.0016"
}

@article{Gendler:2023hwg,
    author = "Gendler, Naomi and Janssen, Oliver and Kleban, Matthew and La Madrid, Joan and Mehta, Viraf M.",
    title = "{Axion minima in string theory}",
    eprint = "2309.01831",
    archivePrefix = "arXiv",
    primaryClass = "hep-th",
    doi = "10.1007/JHEP02(2025)134",
    journal = "JHEP",
    volume = "02",
    pages = "134",
    year = "2025"
}

@article{Choi:2020rgn,
    author = "Choi, Kiwoon and Im, Sang Hui and Sub Shin, Chang",
    title = "{Recent Progress in the Physics of Axions and Axion-Like Particles}",
    eprint = "2012.05029",
    archivePrefix = "arXiv",
    primaryClass = "hep-ph",
    reportNumber = "CTPU-PTC-20-28",
    doi = "10.1146/annurev-nucl-120720-031147",
    journal = "Ann. Rev. Nucl. Part. Sci.",
    volume = "71",
    pages = "225--252",
    year = "2021"
}

@article{Halverson:2019cmy,
    author = "Halverson, James and Long, Cody and Nelson, Brent and Salinas, Gustavo",
    title = "{Towards string theory expectations for photon couplings to axionlike particles}",
    eprint = "1909.05257",
    archivePrefix = "arXiv",
    primaryClass = "hep-th",
    doi = "10.1103/PhysRevD.100.106010",
    journal = "Phys. Rev. D",
    volume = "100",
    number = "10",
    pages = "106010",
    year = "2019"
}

@article{benoit1924note,
  title="{Note sur une m{\'e}thode de r{\'e}solution des {\'e}quations normales provenant de l’application de la m{\'e}thode des moindres carr{\'e}s {\`a} un syst{\`e}me d’{\'e}quations lin{\'e}aires en nombre inf{\'e}rieur {\`a} celui des inconnues (Proc{\'e}d{\'e} du Commandant Cholesky)}",
  author={Benoit, Commandant},
  journal={Bulletin g{\'e}od{\'e}sique},
  volume={2},
  number={1},
  pages={67--77},
  year={1924},
  doi={10.1007/BF03031308}
}

@book{watkins2004fundamentals,
  title="{Fundamentals of Matrix Computations}",
  author={Watkins, David S},
  year={2004},
  publisher={John Wiley \& Sons},
  doi={10.1002/0471249718}
}

@article{Benabou:2026jtv,
    author = "Benabou, Joshua N. and Dainelli, Giulio Alvise and Reig, Mario and Safdi, Benjamin R.",
    title = "{Heterotic String Theory Suggests a QCD Axion Near 0.5 neV}",
    eprint = "2605.04142",
    archivePrefix = "arXiv",
    primaryClass = "hep-th",
    reportNumber = "CERN-TH-2026-098",
    month = "5",
    year = "2026"
}

@article{Peccei:1977ur,
      author         = "Peccei, R. D. and Quinn, Helen R.",
      title          = "{Constraints Imposed by CP Conservation in the Presence
                        of Instantons}",
      journal        = "Phys. Rev.",
      volume         = "D16",
      year           = "1977",
      pages          = "1791-1797",
      doi            = "10.1103/PhysRevD.16.1791",
      reportNumber   = "ITP-572-STANFORD",
      SLACcitation   = "%%CITATION = PHRVA,D16,1791;%%"
}

@article{Peccei:1977hh,
      author         = "Peccei, R. D. and Quinn, Helen R.",
      title          = "{CP Conservation in the Presence of Instantons}",
      journal        = "Phys. Rev. Lett.",
      volume         = "38",
      year           = "1977",
      pages          = "1440-1443",
      doi            = "10.1103/PhysRevLett.38.1440",
      reportNumber   = "ITP-568-STANFORD",
      SLACcitation   = "%%CITATION = PRLTA,38,1440;%%"
}

@article{Wilczek:1977pj,
      author         = "Wilczek, Frank",
      title          = "{Problem of Strong P and T Invariance in the Presence of
                        Instantons}",
      journal        = "Phys. Rev. Lett.",
      volume         = "40",
      year           = "1978",
      pages          = "279-282",
      doi            = "10.1103/PhysRevLett.40.279",
      reportNumber   = "Print-77-0939 (COLUMBIA)",
      SLACcitation   = "%%CITATION = PRLTA,40,279;%%"
}

@article{Weinberg:1977ma,
      author         = "Weinberg, Steven",
      title          = "{A New Light Boson?}",
      journal        = "Phys. Rev. Lett.",
      volume         = "40",
      year           = "1978",
      pages          = "223-226",
      doi            = "10.1103/PhysRevLett.40.223",
      reportNumber   = "HUTP-77/A074",
      SLACcitation   = "%%CITATION = PRLTA,40,223;%%"
}

@article{Preskill:1982cy,
      author         = "Preskill, John and Wise, Mark B. and Wilczek, Frank",
      title          = "{Cosmology of the Invisible Axion}",
      journal        = "Phys. Lett.",
      volume         = "120B",
      year           = "1983",
      pages          = "127-132",
      doi            = "10.1016/0370-2693(83)90637-8",
      reportNumber   = "HUTP-82-A048, NSF-ITP-82-103",
      SLACcitation   = "%%CITATION = PHLTA,120B,127;%%"
}

@article{Dine:1982ah,
      author         = "Dine, Michael and Fischler, Willy",
      title          = "{The Not So Harmless Axion}",
      journal        = "Phys. Lett.",
      volume         = "120B",
      year           = "1983",
      pages          = "137-141",
      doi            = "10.1016/0370-2693(83)90639-1",
      reportNumber   = "UPR-0201T",
      SLACcitation   = "%%CITATION = PHLTA,120B,137;%%"
}

@article{Abbott:1982af,
      author         = "Abbott, L. F. and Sikivie, P.",
      title          = "{A Cosmological Bound on the Invisible Axion}",
      journal        = "Phys. Lett.",
      volume         = "120B",
      year           = "1983",
      pages          = "133-136",
      doi            = "10.1016/0370-2693(83)90638-X",
      reportNumber   = "PRINT-82-0695 (BRANDEIS)",
      SLACcitation   = "%%CITATION = PHLTA,120B,133;%%"
}

@article{Reece:2023iqn,
    author = "Reece, Matthew",
    title = "{Axion-gauge coupling quantization with a twist}",
    eprint = "2309.03939",
    archivePrefix = "arXiv",
    primaryClass = "hep-ph",
    doi = "10.1007/JHEP10(2023)116",
    journal = "JHEP",
    volume = "10",
    pages = "116",
    year = "2023"
}

@article{Choi:2023pdp,
    author = "Choi, Yichul and Forslund, Matthew and Lam, Ho Tat and Shao, Shu-Heng",
    title = "{Quantization of Axion-Gauge Couplings and Noninvertible Higher Symmetries}",
    eprint = "2309.03937",
    archivePrefix = "arXiv",
    primaryClass = "hep-ph",
    reportNumber = "YITP-SB-2023-27, MIT-CTP/5606",
    doi = "10.1103/PhysRevLett.132.121601",
    journal = "Phys. Rev. Lett.",
    volume = "132",
    number = "12",
    pages = "121601",
    year = "2024"
}

@article{Cordova:2023her,
    author = "Cordova, Clay and Hong, Sungwoo and Wang, Lian-Tao",
    title = "{Axion domain walls, small instantons, and non-invertible symmetry breaking}",
    eprint = "2309.05636",
    archivePrefix = "arXiv",
    primaryClass = "hep-ph",
    doi = "10.1007/JHEP05(2024)325",
    journal = "JHEP",
    volume = "05",
    pages = "325",
    year = "2024"
}

@article{Cheng:2025ggf,
    author = "Cheng, Junyi and Gendler, Naomi",
    title = "{Universality in the axiverse}",
    eprint = "2507.12516",
    archivePrefix = "arXiv",
    primaryClass = "hep-th",
    doi = "10.1007/JHEP11(2025)012",
    journal = "JHEP",
    volume = "11",
    pages = "012",
    year = "2025"
}

@article{DiUbaldo:2026rly,
    author = "Di Ubaldo, Gabriele and Iliesiu, Luca V. and Lin, Henry W. and Yan, Cynthia",
    title = "{Positivity of the gravitational path integral implies the axionic weak gravity conjecture}",
    eprint = "2605.05305",
    archivePrefix = "arXiv",
    primaryClass = "hep-th",
    reportNumber = "RIKEN-iTHEMS-Report-26",
    month = "5",
    year = "2026"
}

@article{Maldacena:2026jqd,
    author = "Maldacena, Juan and Maloney, Alexander and McPeak, Brian",
    title = "{Wormholes and the imaginary distance bound}",
    eprint = "2605.05336",
    archivePrefix = "arXiv",
    primaryClass = "hep-th",
    month = "5",
    year = "2026"
}

@article{Long:2016jvd,
    author = "Long, Cody and McAllister, Liam and Stout, John",
    title = "{Systematics of Axion Inflation in Calabi-Yau Hypersurfaces}",
    eprint = "1603.01259",
    archivePrefix = "arXiv",
    primaryClass = "hep-th",
    doi = "10.1007/JHEP02(2017)014",
    journal = "JHEP",
    volume = "02",
    pages = "014",
    year = "2017"
}

@article{Demirtas:2022hqf,
    author = "Demirtas, Mehmet and Rios-Tascon, Andres and McAllister, Liam",
    title = "{CYTools: A Software Package for Analyzing Calabi-Yau Manifolds}",
    eprint = "2211.03823",
    archivePrefix = "arXiv",
    primaryClass = "hep-th",
    month = "11",
    year = "2022"
}

@article{Martucci:2024trp,
    author = "Martucci, Luca and Risso, Nicol{\`o} and Valenti, Alessandro and Vecchi, Luca",
    title = "{Wormholes in the axiverse, and the species scale}",
    eprint = "2404.14489",
    archivePrefix = "arXiv",
    primaryClass = "hep-th",
    doi = "10.1007/JHEP07(2024)240",
    journal = "JHEP",
    volume = "07",
    pages = "240",
    year = "2024"
}

@article{Mehta:2020kwu,
    author = "Mehta, Viraf M. and Demirtas, Mehmet and Long, Cody and Marsh, David J. E. and Mcallister, Liam and Stott, Matthew J.",
    title = "{Superradiance Exclusions in the Landscape of Type IIB String Theory}",
    eprint = "2011.08693",
    archivePrefix = "arXiv",
    primaryClass = "hep-th",
    reportNumber = "KCL-PH-TH/2020-77",
    month = "11",
    year = "2020"
}

@inproceedings{Morrison94,
	Archiveprefix = {arXiv},
	Author = {David R. Morrison},
	Booktitle = {Proceedings of the Hirzebruch 65 Conference on Algebraic Geometry (Ramat Gan, 1993)},
	Eprint = {alg-geom/9407007},
	Pages = {361--376},
	Primaryclass = {alg-geom},
	Title = {{Beyond the K{\"a}hler Cone}},
	Year = {1994}}

@incollection{Morrison93,
	Archiveprefix = {arXiv},
	Author = {Morrison, David R.},
	Booktitle = {Journ\'ees de g\'eom\'etrie alg\'ebrique d'Orsay - Juillet 1992},
	Eprint = {alg-geom/9304007},
	Number = {218},
	Primaryclass = {alg-geom},
	Publisher = {Soci\'et\'e math\'ematique de France},
	Series = {Ast\'erisque},
	Title = {Compactifications of moduli spaces inspired by mirror symmetry},
	Url = {http://www.numdam.org/item/AST_1993__218__243_0/},
	Year = {1993}}

@article{Etheredge:2026rio,
    author = "Etheredge, Muldrow and Reece, Matthew and Rudelius, Tom and Tudball, Christopher",
    title = "{Sharpening the Supersymmetric Axion Weak Gravity Conjecture}",
    eprint = "2605.22912",
    archivePrefix = "arXiv",
    primaryClass = "hep-th",
    reportNumber = "MPP-2026-93",
    month = "5",
    year = "2026"
}
\end{document}